\documentclass[12pt, draftclsnofoot, onecolumn]{IEEEtran}
\usepackage{cite}

\usepackage{hyperref}
\usepackage{booktabs}
\usepackage{multirow}

\ifCLASSINFOpdf
   \usepackage[pdftex]{graphicx}
   \graphicspath{{../pdf/}{../jpeg/}}
   \DeclareGraphicsExtensions{.pdf,.jpeg,.png}
\else
   \usepackage[dvips]{graphicx}
   \graphicspath{{../eps/}}
   \DeclareGraphicsExtensions{.eps}
\fi
\usepackage{amsmath,amssymb}
\usepackage{algorithmic}
\makeatletter
\newcommand{\removelatexerror}{\let\@latex@error\@gobble}
\makeatother
\usepackage{algorithm}
\ifCLASSOPTIONcompsoc
  \usepackage[caption=false,font=normalsize,labelfont=sf,textfont=sf]{subfig}
\else
  \usepackage[caption=false,font=footnotesize]{subfig}
\fi
\begin{document}
%
\title{Low Complexity SLP: An Inversion-Free, Parallelizable ADMM Approach}
%
%
%

\author{Junwen~Yang,
		Ang~Li,~\IEEEmembership{Senior~Member,~IEEE,}
		Xuewen~Liao,~\IEEEmembership{Member,~IEEE,}	
and~Christos~Masouros,~\IEEEmembership{Senior~Member,~IEEE}
\thanks{Manuscript received XXXX; revised XXXX. \emph{(Corresponding author: Xuewen Liao.)}}
\thanks{J. Yang, A. Li are with the School of Information and Communications Engineering, Faculty of Electronic and Information Engineering, Xi'an Jiaotong University, Xi'an, Shaanxi 710049, China (e-mail: jwyang@stu.xjtu.edu.cn; ang.li.2020@xjtu.edu.cn).}
\thanks{X. Liao is with the School of Information and Communications Engineering, Faculty of Electronic and Information Engineering, Xi’an Jiaotong University, Xi’an, Shaanxi 710049, China, and also with the National Mobile Communications Research Laboratory, Southeast University, Nanjing 210096, China (e-mail: yeplos@mail.xjtu.edu.cn).}
\thanks{C. Masouros is with the Department of Electronic and Electrical Engineering, University College London, London WC1E 7JE, U.K. (e-mail: c.masouros@ucl.ac.uk).}
}

\maketitle

\begin{abstract}
We propose a parallel constructive interference (CI)-based symbol-level precoding (SLP) approach for massive connectivity in the downlink of multiuser multiple-input single-output (MU-MISO) systems, with only local channel state information (CSI) used at each processor unit and limited information exchange between processor units. By reformulating the power minimization (PM) SLP problem and exploiting the separability of the corresponding reformulation, the original problem is decomposed into several parallel subproblems via the ADMM framework with closed-form solutions, leading to a substantial reduction in computational complexity. The sufficient condition for guaranteeing the convergence of the proposed approach is derived, based on which an adaptive parameter tuning strategy is proposed to accelerate the convergence rate. To avoid the large-dimension matrix inverse operation, an efficient algorithm is proposed by employing the standard proximal term and by leveraging the singular value decomposition (SVD). Furthermore, a prox-linear proximal term is adopted to fully eliminate the matrix inversion, and a parallel inverse-free SLP (PIF-SLP) algorithm is finally obtained. Numerical results validate our derivations above, and demonstrate that the proposed PIF-SLP algorithm can significantly reduce the computational complexity compared to the state-of-the-arts.
\end{abstract}


%
\IEEEpeerreviewmaketitle

\section{Introduction}
%
%
%
%
%
%

\IEEEPARstart{M}{assive} multiuser multiple-input multiple-output (M-MU-MIMO) has been foreseen as one of the key enablers for future wireless communication systems \cite{rusek2012scaling,sanguinetti2019toward,liu2018massive}, as it has the potential to offer tremendous multiplexing gain and array gain, thereby meeting the boosting requirements of spectral efficiency and energy efficiency \cite{marzetta2010noncooperative,ngo2013energy}. As a fundamental factor that affects system performance, interference plays a central role in reaping the benefits of M-MU-MIMO, and needs to be dealt with carefully. As an effective interference management technique in the downlink, precoding has attracted extensive attention \cite{zheng2003diversity}.

Maximum ratio transmission (MRT) precoding is the simplest strategy that maximizes the received signal-to-noise ratio (SNR) \cite{lo1999maximum}. MRT is devised for noise-limited scenarios, when it comes to interference-limited scenarios, zero-forcing (ZF) precoding is a preferable choice \cite{caire2003achievable}, which employs the channel inversion to eliminate the multiuser interference at the price of augmenting noise. As a regularized form of channel inversion, regularized ZF (RZF) precoding is proposed to alleviate the noise-amplifying effect of ZF \cite{peel2005vector}. The aforementioned linear precoding methods are close to optimal only when the number of transmit antennas is far greater than the number of users \cite{yang2013performance}, because in such case the channels of users are asymptotically orthogonal and favorable propagation can be achieved \cite{rusek2012scaling}. On the other hand, as the number of users keeps increasing, there will be a large spread in the singular value of the channel matrix \cite{peel2005vector}, which will dramatically deteriorate the performance of linear precoding methods.

Except for closed-form linear precoding methods described above, there also exist a number of nonlinear precoding approaches in the literature. Dirty paper coding (DPC) is a capacity-achieving nonlinear precoding method that cancels known interference sequentially leveraging the full channel state information (CSI) \cite{costa1983writing}. Nevertheless, the unrealistic assumption of infinite codebook length hinders the practical implementation of DPC. An interference cancellation alternative is the Tomlinson-Harashima precoding (THP), which imposes an integer offset at the transmitter, and a modulo operation is required for the received signal \cite{windpassinger2004precoding}. Despite the near-capacity performance of THP, its encoding and decoding are of great complexity. Instead of sequentially calculating the offset in THP, another nonlinear precoding, the vector perturbation (VP) precoding, jointly selects a perturbation vector via the sphere encoding algorithm and transmits the perturbed signals to the users, which is shown to achieve superior performance and requires a relatively simpler decoding procedure than THP \cite{hochwald2005vector}.

In addition to devising the precoder heuristically or analytically, pursuing the optimal precoding strategy naturally resorts to optimization. For example, the SINR-constrained power minimization (PM) problem aims to minimize the total transmit power, subject to the received SINR target for each user \cite{visotsky1999optimum}. This problem can be solved via the uplink-downlink duality \cite{visotsky1999optimum} or conic programming \cite{wiesel2005linear}, and also can be reformulated into a semidefinite optimization, for which the semidefinite relaxation approach is viable \cite{bengtsson1999optimal}. The inverse problem of PM is the max-min SINR balancing (SB) problem, which maximizes the minimum SINR subject to a total transmit power constraint \cite{palomar2003joint,schubert2004solution,wiesel2005linear}. The SB problem is nonconvex and cannot be reformulated to a convex form. Stimulated by the relationship between SB and PM, iterative algorithms that solve a series of PM problems in a bi-section search have been proposed in the literature to solve the SB problem \cite{wiesel2005linear,karipidis2008quality}. The nonlinear precoding methods are also known as symbol-level precoding (SLP), because their precoding matrices are jointly determined by the CSI and data symbols, and generally redesigned for each symbol slot. From a statistical perspective, the interference is uncontrollable and performs as a deterioration factor, and thereby BLP methods aim to mitigate or eliminate interference. On the other hand, from an instantaneous view, interference is controllable and can be manipulated to enhance signal detection by means of SLP. This was first discussed by the constructive interference (CI) precoding in the context of pre-decorrelation and Pre-Rake \cite{masouros2007novel}. The same concept was introduced to ZF precoding later in \cite{masouros2009dynamic}. As a step further, a correlation rotation precoding technique was designed to rotate both CI and destructive interference (DI) such that the phase of interference is aligned to the signal of interest, based on which DI can be transferred into CI \cite{masouros2010correlation}. The first work to combine CI precoding with optimization was proposed in the context of  VP precoding with limited feedback. Standing on the concept of CI, the optimization-based PM-SLP and SB-SLP schemes were further studied, where the resulting interference is no longer strictly aligned to the signal of interest, but constrained by the CI regions \cite{masouros2015exploiting}, which provides further performance improvements. At the same time, this means that the SLP methods must solve a constrained optimization problem at each symbol slot to obtain the full benefits offered by CI-SLP, resulting substantial computational complexity, especially in M-MU-MIMO settings.

Towards low-complexity and low-latency CI-SLP solutions, plenty of works have endeavored to find efficient and practical SLP solutions. For PM-SLP, the virtual multicast formulation is widely used, by which the optimization variable is shifted from the large-dimension precoding matrix to the small-dimension precoded signal vector \cite{masouros2015exploiting,haqiqatnejad2018power,haqiqatnejad2019approximate}. Subsequently, Lagrange duality is applied to inspect the reformulated problem, whose Lagrangian dual is identified as a nonnegative least-squares (NNLS) problem. A gradient projection algorithm with line search was proposed to solve the NNLS problem \cite{masouros2015exploiting}. With further inspection, the structure of the optimal solution for PM-SLP was analyzed via the Karush–Kuhn–Tucker (KKT) optimality conditions, which leads to a closed-form suboptimal solution for the NNLS problem \cite{haqiqatnejad2018power}. To improve the approximation performance of the suboptimal solution, its improved alternative with an extra validation step was proposed \cite{haqiqatnejad2019approximate}. For SB-SLP, its Lagrangian dual was shown to be a QP optimization over a probability simplex, and the optimal structure of the precoding matrix was derived, based on which a closed-form iterative algorithm with conditional optimality was developed \cite{li2018interference,li2020interference}. On the other hand, deep learning-based low-complexity SLP frameworks, such as CI-NN and SLP-DNet were also proposed\cite{9357329,mohammad2021unsupervised}.

Based on the above descriptions, it can be summarized that most of the existing CI-based precoding approaches need sequential and centralized implementations, while the closed-form suboptimal solutions suffer from performance losses. More importantly, the resource-demanding matrix inverse operation is commonly required, so the resulting complexity of SLP is still high, which hinders its practical implementation. Motivated by these findings, in this paper, we propose a parallel inverse-free SLP (PIF-SLP) scheme for M-MU-MIMO downlink. The main contributions of the paper are summarized as follows.

\begin{enumerate}
\item We propose a parallel CI-SLP approach based on the proximal Jacobian alternating direction method of multipliers (PJ-ADMM) for the PM-SLP problem. For the first time in literature, we take advantage of the separable structure of the PM-SLP problem, and by transferring the inequality constraints into equality constraints with the introduced slack variable vector, the original problem is formulated into an unconstrained problem using the augmented Lagrangian method (ALM). The PJ-ADMM framework is adopted to decouple the unconstrained problem into a series of parallel subproblems, and closed-form solutions are obtained for each subproblem. 

\item We analyze the convergence performance of the proposed parallel CI-SLP approach, and derive the sufficient condition for convergence, which indicates that the parallel SLP approach is guaranteed to converge to the global optimum as long as the proximal term is chosen sufficiently large. However, a larger proximal term will result in a slower convergence rate. Accordingly, an adaptive parameter tuning strategy is developed to speed up the convergence.

\item Based on the above, we propose the PIF-SLP algorithm by adapting a prox-linear proximal term to avoid the matrix inverse operation for further complexity reduction. Specifically, the Hessian of the quadratic penalty term in the Lagrangian function is approximated with an identical proximal matrix, and hence the corresponding matrix inversion can be replaced by scalar division. Meanwhile, the required number of matrix multiplication is also reduced.

\item We further propose a low-coordination overhead decentralized scheme to alleviate the coordination overhead, at the cost of slightly increased computational overhead. By rearranging the closed-form solutions of the subproblems and incorporating the updates of global variables in the parallel processor units, the extra consensus node is removed, which reduces the coordination overhead. The computational complexity and the coordination overhead between processor units of the parallel SLP approach are also studied analytically.
\end{enumerate}

Monte Carlo simulations are conducted to validate our analysis as well as the effectiveness of the proposed schemes, where it is demonstrated that the proposed PIF-SLP algorithm can greatly reduce the computational burden of the CI-SLP without performance loss. A scalable complexity-performance trade-off of the parallel SLP approach is also observed.

The remainder of this paper is organized as follows. Section \ref{secModel} introduces the system model and CI, as well as the canonical PM-SLP problem formulation. Section \ref{ADMM} reformulates the canonical problem based on separability and slackness, where ALM and ADMM are further introduced. The proposed parallel SLP approach and its sufficient condition of convergence are presented in section \ref{PJADMM}, including the adaptive parameter tuning strategy and the final PIF-SLP algorithm. Section \ref{complexity} provides the computational complexity and coordination overhead analysis. Numerical results are presented in Section \ref{results}, and Section \ref{conclusion} concludes the paper.

\textbf{Notation:} Scalars, vectors, and matrices, are represented by plain lowercase, boldface lowercase, and boldface capital letters, respectively. $(\cdot)^T$, $(\cdot)^H$, and $(\cdot)^{-1}$ denote transpose, conjugate transpose, and inverse operators, respectively. $\mathbb{C}^{M\times N}$ and $\mathbb{R}^{M\times N}$ denote the sets of $M\times N$ matrices with complex and real entries, respectively. $\left| \cdot \right|$ represents the absolute value of a real scalar or the modulus of a complex scalar. $\left\|\cdot\right\|$ denotes the Euclidean norm of a vector or spectral norm of a matrix. $\Re \{\cdot\}$ and $\Im \{\cdot\}$ respectively denote the real part and imaginary part of a complex input. $\succeq$ denotes element-wise inequality. $\mathbf{0}$, $\mathbf{1}$, and $\mathbf{I}$ represent respectively, the all-zeros vector, the all-ones vector, and the identity matrix with appropriate dimensions. $\max\{\cdot\}$ represents the elementwise maximum. $\oslash$ denotes the element-wise division. $diag\{\cdot\}$ returns a vector consisting of the main diagonal elements of a input matrix.

\section{System Model and Problem Formulation}
\label{secModel}
\subsection{System Model}
\label{subsecModel}
We consider a downlink massive MU-MISO system, where a base station (BS) equipped with $N_t$ antennas serves $K$ single-antenna users in the same time-frequency resource. The modulated data symbol vector $\tilde{\mathbf{s}}\triangleq [\tilde{s}_{1},\cdots , \tilde{s}_{K}]^{T}\in\mathbb{C}^{K}$ is composed of $K$ independent symbols randomly drawn from a normalized $\mathcal{M}$-ary PSK constellation, which is mapped to the transmit signal $\tilde{\mathbf{x}}\triangleq [\tilde{x}_{1},\cdots , \tilde{x}_{N_{t}}]^{T}\in\mathbb{C}^{N_{t}}$ at the BS via SLP. The received signal of user $k$ is expressed as
\begin{IEEEeqnarray}{rCl}
y_{k}&=&\tilde{\mathbf{h}}^T_{k}\tilde{\mathbf{x}}+n_{k} ,
\end{IEEEeqnarray}
where $\tilde{\mathbf{h}}_{k}\in\mathbb{C}^{N_t}$ denotes the quasi-static Rayleigh flat-fading channel vector between BS and user $k$, and $n_k$ is the circularly symmetric complex zero-mean Gaussian white noise with variance $\sigma^2_k$ at user $k$. The above signal model can be written in a more compact form as
\begin{IEEEeqnarray}{rCl}
\mathbf{y}=\tilde{\mathbf{H}}\tilde{\mathbf{x}}+\mathbf{n},
\end{IEEEeqnarray}
where $\mathbf{y}\triangleq[y_{1},\cdots , y_{K}]^{T}\in\mathbb{C}^K$ and $\mathbf{n}\triangleq[n_{1},\cdots , n_{K}]^{T}\in\mathbb{C}^K$ denote the received signal and noise at all $K$ users, respectively. $\tilde{\mathbf{H}}\triangleq[\tilde{\mathbf{h}}_{1},\cdots,\tilde{\mathbf{h}}_{K}]^T\in\mathbb{C}^{K\times N_{t}}$ is the channel matrix. To focus on the precoding design, perfect CSI is assumed throughout this paper.

\subsection{Constructive Interference}
\begin{figure}[!t]
\centering
\includegraphics{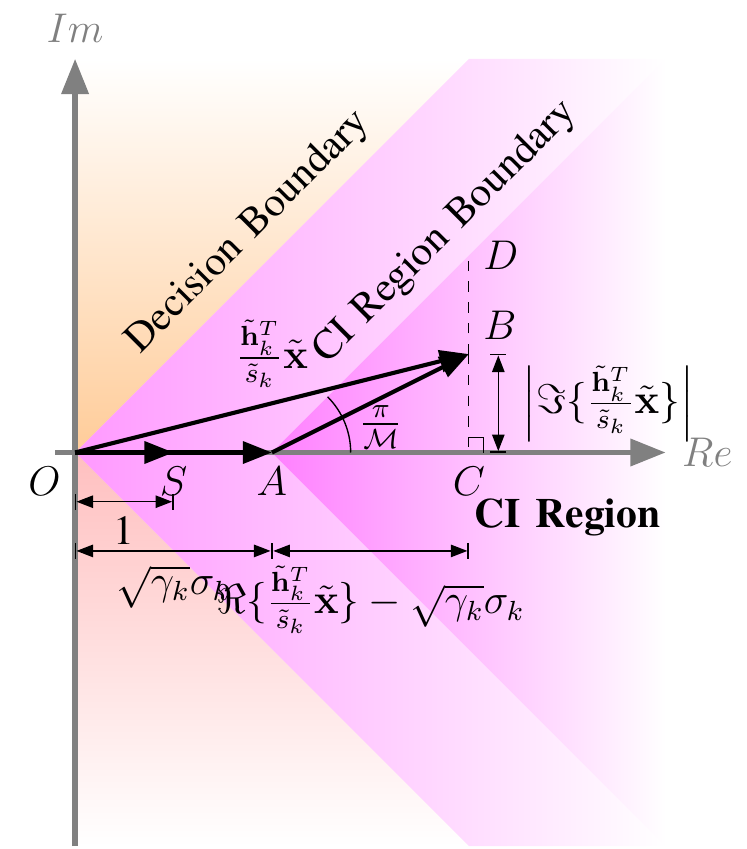}
\caption{Illustration of CI regions for a generic $\mathcal{M}$-PSK modulation.}
\label{fig_CI}
\end{figure}

CI precoding was first introduced in \cite{masouros2007novel}, which reveals that the constructive and destructive interference pattern of the noiseless received signal $\left\{\tilde{\mathbf{h}}^T_{k}\tilde{\mathbf{x}}\right\}$ is jointly determined by CSI and data symbols. Based on this fact, interference can be predicted and further exploited using SLP, which judiciously utilizes CSI and data symbols to optimize the transmit signal, such that all the multiuser interference add up constructively at receivers \cite{li2020tutorial}. Therefore, the received instantaneous SINR at user $k$ is given as
$\mathrm{SINR}_{k}\triangleq\frac{\left|\tilde{\mathbf{h}}^T_{k}\tilde{\mathbf{x}}\right|^2}{\sigma^2_k}$.
Since all interference is exploited via SLP, the SINR is equivalent to the conventional signal-to-noise ratio (SNR).

Geometrically, CI is achieved as long as the noiseless received signal of each user lies in the symbol-specified CI region in the complex plane, where the CI region refers to a polyhedron bounded by hyperplanes parallel to decision boundaries or Voronoi edges \cite{masouros2015exploiting,haqiqatnejad2018constructive}, and the only vertex of one CI region is the SINR threshold-dependent nominal constellation symbol, as depicted in Fig. \ref{fig_CI}. Without loss of generality, let $\tilde{s}_k$ be the symbol of interest for user $k$, which is an arbitrary constellation point drawn from a normalized $\mathcal{M}$-PSK constellation as described in Section \ref{subsecModel}. We rotate $\tilde{s}_k$ to the positive real axis, thereby the rotated symbol is 1, which corresponds to $\overrightarrow{OS}$ in Fig. \ref{fig_CI}. Other related signals are rotated by the same phase. Consequently, the received noiseless signal of user $k$, $\tilde{\mathbf{h}}^T_{k}\tilde{\mathbf{x}}$, turns out to $\frac{\tilde{\mathbf{h}}^T_{k}}{\tilde{s}_k}\tilde{\mathbf{x}}$, which is denoted by $\overrightarrow{OB}$ in Fig. \ref{fig_CI}. For a given instantaneous SINR threshold $\gamma_k$ for user $k$, the nominal constellation point is equivalent to $\sqrt{\gamma_k}\sigma_k\tilde{s}_k$
. We introduce $\overrightarrow{OA}$ as the rotated nominal constellation point, which is also the only vertex of the interested CI region. When $\overrightarrow{OB}$ is located in the depicted CI region, the received signal is pushed away from decision boundaries and the instantaneous SINR is guaranteed to be no less than the prescribed threshold $\gamma_k$. One of the criteria that specifies the location of $\overrightarrow{OB}$ in the CI region is $\left|\overrightarrow{CD}\right|\geq \left|\overrightarrow{CB}\right|$. Accordingly, the corresponding explicit mathematical formulation of CI constraints for $\mathcal{M}$-PSK signaling can be written as
\begin{IEEEeqnarray}{rCl}
\Re\left\{\hat{\mathbf{h}}^T_{k}\tilde{\mathbf{x}}\right\}-\frac{\left|\Im\left\{\hat{\mathbf{h}}^T_{k}\tilde{\mathbf{x}}\right\}\right|}{\tan \frac{\pi}{\mathcal{M}}}\geq\sqrt{\gamma_k}\sigma_k,\, \forall k,
\end{IEEEeqnarray}
where $\hat{\mathbf{h}}^T_{k}\triangleq\frac{\tilde{\mathbf{h}}^T_{k}}{\tilde{s}_k}$, $\gamma_k$ denotes the pre-defined instantaneous SINR threshold for user $k$. It is worth noting that the CI constraint for each user already incorporates the SINR constraint.\footnote{The CI constraints can be readily extended to multi-level modulation such as QAM by employing the symbol-scaling metric \cite{masouros2014vector}.}

\subsection{SLP for Power Minimization}
Throughout this paper, we are interested in minimizing the total transmit power subject to CI constraints, which is known as the PM-SLP problem. This optimization problem can be formulated as
\begin{IEEEeqnarray}{rCl}
\label{eq_originalPM}
\begin{IEEEeqnarraybox}[][c]{rCl}
\mathcal{P}_{1}&: &\min_{\tilde{\mathbf{x}}} \left\|\tilde{\mathbf{x}}\right\|^2\\
&&\mathrm{s.t.}\, \Re\left\{\hat{\mathbf{h}}^T_{k}\tilde{\mathbf{x}}\right\}-\frac{\left|\Im\left\{\hat{\mathbf{h}}^T_{k}\tilde{\mathbf{x}}\right\}\right|}{\tan \frac{\pi}{\mathcal{M}}}\geq\sqrt{\gamma_k}\sigma_k,\, \forall k.
\end{IEEEeqnarraybox}
\end{IEEEeqnarray}
The quadratic objective function and linear constraints indicate that this problem is convex, and hence can be handled via off-the-shelf solvers. Unfortunately, most generic solvers, e.g., SeDuMi and SDPT3, are based on the high-complexity interior-point method (IPM). To alleviate the computational burden, efficient algorithms based on gradient projection method \cite{masouros2015exploiting}, suboptimal closed-form solution \cite{haqiqatnejad2018power}, and improved suboptimal closed-form solution \cite{haqiqatnejad2019approximate} were proposed. Existing works, however, focus on centralized iterative algorithms and ignore the separable nature of the PM-SLP problem. By exploiting such separability, we propose a parallel CI-SLP precoding approach in this paper based on the PJ-ADMM, as shown below.

\section{ALM and Conventional ADMM for PM-SLP}
\label{ADMM}
In this section, we investigate the structure of the PM-SLP optimization problem and reveal its separable nature. ALM is used to tackle the reformulated problem subsequently. Conventional Gauss-Seidel ADMM and Jacobian ADMM are further employed to exploit the separability to arrive at sequential and parallel solutions. In the next section, we present the proposed PIF-SLP approach.
\subsection{Separability and Slackness}
The real-valued equivalence of $\mathcal{P}_1$ can be written as
\begin{IEEEeqnarray}{rCl}
\begin{IEEEeqnarraybox}[][c]{rCl}
\mathcal{P}_{2}&: &\min_{\mathbf{x}} \left\|\mathbf{x}\right\|^2\\
&&\mathrm{s.t.}\, \mathbf{N}\mathbf{S}_{k}\mathbf{H}_{k}\mathbf{x}\succeq\sqrt{\gamma_k}\sigma_k\mathbf{1},\, \forall k,
\end{IEEEeqnarraybox}
\end{IEEEeqnarray}
where $\mathbf{x}\triangleq{\left[\begin{array}{c}\Re \left\{\tilde{\mathbf{x}}\right\} \\
\Im \left\{\tilde{\mathbf{x}}\right\} \end{array}\right]}\in\mathbb{R}^{2N_t}$, $
\mathbf{N}\triangleq{\left[\begin{array}{cc}1 & -\frac{1}{\tan\frac{\pi}{\mathcal{M}}} \\
1 & \frac{1}{\tan\frac{\pi}{\mathcal{M}}} \end{array}\right]}\in\mathbb{R}^{2\times 2}$, $\mathbf{S}_{k}\triangleq{\left[\begin{IEEEeqnarraybox*}[][c]{,c/c,}
\Re \left\{\frac{1}{\tilde {s}_k}\right\} & -\Im \left\{\frac{1}{\tilde {s}_k}\right\} \\
\Im \left\{\frac{1}{\tilde {s}_k}\right\} & \Re \left\{\frac{1}{\tilde {s}_k}\right\} \end{IEEEeqnarraybox*}\right]}\in\mathbb{R}^{2\times 2}$, and $\mathbf{H}_{k}\triangleq{\left[\begin{IEEEeqnarraybox*}[][c]{,c/c,}
\Re \left\{\tilde{\mathbf {h}}^T_k\right\} & -\Im \left\{\tilde{\mathbf{h}}^T_k\right\} \\ 
\Im \left\{\tilde{\mathbf {h}}^T_k\right\} & \Re \left\{\tilde{\mathbf {h}}^T_k\right\} \end{IEEEeqnarraybox*}\right]}\in\mathbb{R}^{2\times 2N_t}$. We further introduce $\bar{\mathbf{A}}_k\triangleq\mathbf{N}\mathbf{S}_{k}\mathbf{H}_{k}$, and $\mathbf{b}_k\triangleq\sqrt{\gamma_k}\sigma_k\mathbf{1}$. Accordingly, the CI constraints become
\begin{IEEEeqnarray}{rCl}
\bar{\mathbf{A}}_{k}\mathbf{x}\succeq\mathbf{b}_k,\, \forall k.
\end{IEEEeqnarray}
Stacking the CI constraints, the compact formulation can be written as
\begin{IEEEeqnarray}{rCl}
\label{inequality constraint}
\mathbf{A}\mathbf{x}\succeq\mathbf{b},
\end{IEEEeqnarray}
where $\mathbf{A}\triangleq\left[\bar{\mathbf{A}}^T_1,\cdots,\bar{\mathbf{A}}^T_K\right]^T\in\mathbb{R}^{2K\times 2N_t}$, $\mathbf{b}\triangleq\left[\mathbf{b}^T_1,\cdots,\mathbf{b}^T_K\right]^T\in\mathbb{R}^{2K}$. We can identify that the left-hand side of (\ref{inequality constraint}) can be expressed as a linear combination of the columns of $\mathbf{A}$, i.e., $\sum_{i=1}^{2N_t}\mathbf{a}_{i}x_i$, where $\mathbf{a}_i$ is the $i$-th column of $\mathbf{A}$, $x_i$ is the $i$-th entry of $\mathbf{x}$. Accordingly, $\mathcal{P}_2$ can be rearranged as
\begin{IEEEeqnarray}{rCl}
\begin{IEEEeqnarraybox}[][c]{rCl}
\label{eq_serarablePM}
\mathcal{P}_{3}&: &\min_{\mathbf{x}_i} \sum_{i=1}^{N}\left\|\mathbf{x}_i\right\|^2\\
&&\mathrm{s.t.}\, \sum_{i=1}^{N}\mathbf{A}_{i}\mathbf{x}_i\succeq\mathbf{b},
\end{IEEEeqnarraybox}
\end{IEEEeqnarray}
where $\mathbf{x}_i\in\mathbb{R}^{n_i}$ with $\sum^N_{i=1}n_i=2N_t$ is the $i$-th block of $\mathbf{x}$, composed of the adjacent and/or disadjacent elements of $\mathbf{x}$, and $\mathbf{A}_i\in\mathbb{R}^{2K\times n_i}$ is the $i$-th column block of $\mathbf{A}$, each column of which is uniquely taken from the columns of $\mathbf{A}$. Mathematically, for the adjacent case, $\mathbf{x}=\left[\mathbf{x}^T_1,\cdots,\mathbf{x}^T_N\right]^T$, $\mathbf{A}=\left[\mathbf{A}_1,\cdots,\mathbf{A}_N\right]$, while for the disadjacent case, $\mathbf{x}_i=\mathbf{E}^T_i\mathbf{x}$, $\mathbf{A}_i=\mathbf{A}\mathbf{E}_i$, where $\mathbf{E}_i\in\mathbb{R}^{2N_t\times n_i}$, and each column of $\left\{\mathbf{E}_i\right\}$ is uniquely picked from the columns of the $2N_t\times 2N_t$ identity matrix. With such formulation, $\mathcal{P}_3$ is partitioned into $N$ blocks, here we do not confine the number of blocks, so long as $N$ is a positive integer not greater than $2N_t$.

We reformulate $\mathcal{P}_3$ by introducing a slack variable vector $\mathbf{c}\in\mathbb{R}^{2K}_+$ to replace the original inequality constraints as follows:
\begin{IEEEeqnarray}{rCl}
\begin{IEEEeqnarraybox}[][c]{rCl}
\mathcal{P}_{4}&: &\min_{\mathbf{x}_i,\mathbf{c}} \sum_{i=1}^{N}\left\|\mathbf{x}_i\right\|^2\\
&&\mathrm{s.t.}\, \sum_{i=1}^{N}\mathbf{A}_{i}\mathbf{x}_i=\mathbf{b}+\mathbf{c},\\
&&\qquad\mathbf{c}\succeq 0.
\end{IEEEeqnarraybox}
\end{IEEEeqnarray}
Since the feasible region of the slack variable $\mathbf{c}$ is $\mathbb{R}^{2K}_+$, by introducing an indicator function, the nonnegativity constraints can be incorporated into the objective function:
\begin{IEEEeqnarray}{rCl}
\begin{IEEEeqnarraybox}[][c]{rCl}
\mathcal{P}_{5}&: &\min_{\mathbf{x}_i,\mathbf{c}} \sum_{i=1}^{N}\left\|\mathbf{x}_i\right\|^2+\mathcal{I}_{\mathbb{R}^{2K}_+}\left(\mathbf{c}\right)\\
&&\mathrm{s.t.}\, -\sum_{i=1}^{N}\mathbf{A}_{i}\mathbf{x}_i+\mathbf{b}+\mathbf{c}=\mathbf{0},
\end{IEEEeqnarraybox}
\end{IEEEeqnarray}
where $\mathcal{I}_{\mathbb{R}^{2K}_+}$ is the indicator function of $\mathbb{R}^{2K}_+$ given by
\begin{IEEEeqnarray}{rCl}
\mathcal{I}_{\mathbb{R}^{2K}_+}\left(\mathbf{c}\right)=
\begin{cases}
0,&\text{if } \mathbf{c}\in\mathbb{R}^{2K}_+,\\
+\infty, &\text{otherwise.}
\end{cases}
\end{IEEEeqnarray}

\subsection{ALM}
The corresponding augmented Lagrangian function of $\mathcal{P}_5$ is given by
\begin{IEEEeqnarray}{rCl}
\label{augmented Lagrangian}
\mathcal{L}_\rho\left(\mathbf{x},\mathbf{c},\boldsymbol{\lambda}\right)&=&\sum_{i=1}^{N}\left\|\mathbf{x}_i\right\|^2+I_{\mathbb{R}^{2K}_+}\left(\mathbf{c}\right)+\boldsymbol{\lambda}^T\left(-\sum_{i=1}^{N}\mathbf{A}_{i}\mathbf{x}_i+\mathbf{b}+\mathbf{c}\right)+\frac{\rho}{2}\left\|-\sum_{i=1}^{N}\mathbf{A}_{i}\mathbf{x}_i+\mathbf{b}+\mathbf{c}\right\|^2\nonumber
\\
&= &\sum_{i=1}^{N}\left\|\mathbf{x}_i\right\|^2+I_{\mathbb{R}^{2K}_+}\left(\mathbf{c}\right)+\frac{\rho}{2}\left\|-\sum_{i=1}^{N}\mathbf{A}_{i}\mathbf{x}_i+\mathbf{b}+\mathbf{c}+\frac{\boldsymbol{\lambda}}{\rho}\right\|^2-\frac{1}{2\rho}\left\|\boldsymbol{\lambda}\right\|^2
\end{IEEEeqnarray}
where $\boldsymbol{\lambda}\in\mathbb{R}^{2K}_+$ is the Lagrange multiplier vector, $\rho$ is a positive penalty parameter that determines the severity of the quadratic penalty on constraint violations. When the value of $\boldsymbol{\lambda}$ is close to the optimal Lagrange multiplier, or the penalty parameter $\rho$ is large, the optimal transmit signal vector $\mathbf{x}$ of $\mathcal{P}_5$ can be well approximated by the unconstrained minima of the augmented Lagrangian \cite{bertsekas2014constrained}. Therefore, the original PM-SLP problem can be solved via the augmented Lagrangian method (ALM). 

Starting with an arbitrary $\boldsymbol{\lambda}^0$, the ALM aims to update the multiplier vector iteratively to approximate the optimal dual solution. A common choice of such approximation is the following gradient iteration:
\begin{IEEEeqnarray}{rCl}
\boldsymbol{\lambda}^{t+1}&=&\boldsymbol{\lambda}^{t}+\rho^t\left(-\mathbf{A}\mathbf{x}^{t+1}+\mathbf{b}+\mathbf{c}^{t+1}\right),
\end{IEEEeqnarray}
where the superscript denotes the iteration index, and $\left(\mathbf{c}^{t+1},\mathbf{x}^{t+1}\right)$ is any vector that minimizes $\mathcal{L}_{\rho^t}\left(\mathbf{x},\mathbf{c},\boldsymbol{\lambda}^t\right)$, namely,
\begin{IEEEeqnarray}{rCl}
\left(\mathbf{c}^{t+1},\mathbf{x}^{t+1}\right)&=&\arg\min_{\mathbf{c},\mathbf{x}}\mathcal{L}_{\rho^t}\left(\mathbf{x},\mathbf{c},\boldsymbol{\lambda}^t\right).
\end{IEEEeqnarray}

The standard ALM guarantees global convergence with a theoretical linear convergence rate, while its convergence rate is in general faster in practical problems. Despite the promising convergence performance, the ALM involves a joint optimization of primal variables, and hence can not take advantage of the separability.
\subsection{Gauss-Seidel ADMM}
\label{subsec_ADMM}
Given the current iteration variables $\left( \mathbf{c}^{t},\mathbf{x}^{t},\boldsymbol{\lambda}^{t}\right)$, the ADMM generates new iteration variables $\left( \mathbf{c}^{t+1},\mathbf{x}^{t+1},\boldsymbol{\lambda}^{t+1}\right)$  via alternating optimization. Applying standard ADMM to the separable PM-SLP problem, $\left( \mathbf{c}^{t+1},\mathbf{x}^{t+1}, \boldsymbol{\lambda}^{t+1}\right)$ is updated via the following steps:
\begin{IEEEeqnarray}{rCl}
\IEEEyesnumber\IEEEyessubnumber*
\label{eq_Gauss-Seidel_a}
\mathbf{c}^{t+1}&=&\arg \min_{\mathbf{c}}\mathcal{L}_\rho\left(\mathbf{x}^{t}_1,\cdots,\mathbf{x}^{t}_N,\mathbf{c},\boldsymbol{\lambda}^{t}\right),
\\
\label{eq_Gauss-Seidel_b}
\mathbf{x}^{t+1}_i&=&\arg \min_{\mathbf{x}_i}\mathcal{L}_\rho\left(\mathbf{x}^{t+1}_{<i},\mathbf{x}_i,\mathbf{x}^{t}_{>i},\mathbf{c}^{t+1},\boldsymbol{\lambda}^{t}\right),\forall i,
\\
\label{eq_Gauss-Seidel_c}
\boldsymbol{\lambda}^{t+1}&=&\boldsymbol{\lambda}^{t}+\rho\left(-\sum_{i=1}^{N}\mathbf{A}_{i}\mathbf{x}^{t+1}_i+\mathbf{b}+\mathbf{c}^{t+1}\right).
\end{IEEEeqnarray}

We can observe that the transmit signal $\mathbf{x}^{t+1}_i$ is calculated by a sweep of Gauss-Seidel updates, namely, $\mathbf{x}^{t+1}_i$ is sequentially updated one after another. While the direct extended Gauss-Seidel ADMM does not necessarily converge for $N\geq3$ \cite{chen2016direct}, it is still efficient at solving many practical problems. Thanks to the strong convexity of the objective function of PM-SLP, Gauss-Seidel ADMM is applicable. On the other hand, however, although Gauss-Seidel ADMM is able to partition the original PM-SLP problem into several subproblems and allow distributed processing, parallel processing is still not achievable. Therefore, Gauss-Seidel ADMM is inefficient for large-scale MIMO precoding.

\subsection{Jacobian ADMM}
To enable parallel processing, Jacobian ADMM can be adopted, which minimizes the augmented Lagrangian with respect to $\mathbf{x}_1,\cdots,\mathbf{x}_N$ in a parallel fashion, while keeping the updates of the remaining variables unchanged, given by
\begin{IEEEeqnarray}{rrCl}
\label{eq_JADMM_b}
&\mathbf{x}^{t+1}_i&=&\arg \min_{\mathbf{x}_i}\mathcal{L}_\rho\left(\mathbf{x}_i,\mathbf{x}^{t}_{\neq i},\mathbf{c}^{t+1},\boldsymbol{\lambda}^{t}\right),\forall i.
\end{IEEEeqnarray}
The above full decomposition and parallelization, however, is achieved at the expense of a degraded practical convergence performance compared to the Gauss-Seidel ADMM. It was shown in \cite{he2015full} that the Jacobian ADMM iterations may be divergent, thus the output of the preceding Jacobian updates may not be used in the next iteration directly.

To design a Jacobian ADMM with guaranteed convergence, \cite{he2015full} suggests inserting an underrelaxation step between every two adjacent Jacobian ADMM iterates, given by
\begin{IEEEeqnarray}{rCl}
\label{eq_underrelaxation}
\mathbf{u}^{t+1}=\mathbf{u}^{t}-\alpha\left(\mathbf{u}^{t}-\bar{\mathbf{u}}^{t}\right),
\end{IEEEeqnarray}
where $\mathbf{u}^{t}\triangleq\left(\mathbf{x}^{t}_1,\mathbf{x}^{t}_2\cdots,\mathbf{x}^{t}_N,\boldsymbol{\lambda}^{t}\right)$, $\bar{\mathbf{u}}^{t}\triangleq\left(\bar{\mathbf{x}}^{t}_1,\bar{\mathbf{x}}^{t}_2\cdots,\bar{\mathbf{x}}^{t}_N,\bar{\boldsymbol{\lambda}}^{t}\right)$ denotes the output of the original Jacobian ADMM with the input $\mathbf{u}^{t}$, $\alpha>0$ is a chosen step size. Note that, as an exactly updated intermediate variable, the slack variable $\mathbf{c}$ is excluded from $\mathbf{u}$. The motivation for the above underrelaxation step lies in the fact that the Jacobian decomposition may have poor accuracy to approximate the joint optimization step of ALM, and \cite{he2015full} proposes to compensate for the accuracy loss by combining the last iterate $\mathbf{u}^{t}$ with $\bar{\mathbf{u}}^{t}$ approximately in aid of the step size $\alpha$. Its worst-case $O(1/t)$ convergence rate measured by the iteration complexity in both the ergodic and nonergodic senses is established.

Another way to enhance convergence of the Jacobian ADMM is to regularize each decomposed problem by a proximal term \cite{he2016proximal}. By adopting a proximal Jacobian decomposition method of ALM, the iteration steps are given by
\begin{IEEEeqnarray}{rCl}
\IEEEyesnumber\IEEEyessubnumber*
\label{eq_PJ-ALM_a}
\mathbf{c}^{t+1}&=&\arg \min_{\mathbf{c}}\mathcal{L}_\rho\left(\mathbf{x}^{t}_1,\cdots,\mathbf{x}^{t}_N,\mathbf{c},\boldsymbol{\lambda}^{t}\right),
\\
\label{eq_PJ-ALM_b}
\mathbf{x}^{t+1}_i&=&\arg \min_{\mathbf{x}_i}\mathcal{L}_\rho\left(\mathbf{x}^{t}_{\neq i},\mathbf{x}_i,\mathbf{c}^{t+1},\boldsymbol{\lambda}^{t}\right)+\frac{\tau\rho}{2}\left\|\mathbf{A}_i\left(\mathbf{x}_i-\mathbf{x}^t_i\right)\right\|^2,\forall i,
\\
\label{eq_PJ-ALM_c}
\boldsymbol{\lambda}^{t+1}&=&\boldsymbol{\lambda}^{t}+\rho\left(-\sum_{i=1}^{N}\mathbf{A}_{i}\mathbf{x}^{t+1}_i+\mathbf{b}+\mathbf{c}^{t+1}\right),
\end{IEEEeqnarray}
where $\tau>0$ is a proximal coefficient that controls the proximity of the new iterate to the last one. It is essentially one type of proximal Jacobian ADMM, but to discriminate it from the PJ-ADMM in the next section and follow the terminology in \cite{he2016proximal}, we refer it to PJ-ALM. It was shown in \cite{he2016proximal} that if the proximal coefficient is sufficiently large, i.e., $\tau\geq N-1$, the convergence of the proximal Jacobian decomposition of ALM can be guaranteed.

\section{Proposed PIF-SLP Approach}
\label{PJADMM}
In the previous section, we have revealed the separability of the original PM-SLP optimization problem (\ref{eq_originalPM}) by inspecting and rearranging its structure to facilitate distributed and parallel processing. In this section, we adopt a more general and flexible PJ-ADMM framework in \cite{deng2017parallel} to solve the reformulated PM-SLP problem with closed-form solutions for each subproblem. PJ-ADMM is similar to the preceding relaxation and regularization idea, while its relaxation step is split into primal and dual relaxation, where the primal relaxation is replaced by a flexible quadratic proximal regularization term for each subproblem. The PJ-ADMM procedure for PM-SLP is formulated as
\begin{IEEEeqnarray}{rCl}
\label{eq_PJADMM}
\IEEEyesnumber\IEEEyessubnumber*
\mathbf{c}^{t+1}&=&\arg \min_{\mathbf{c}}\mathcal{L}_\rho\left(\mathbf{x}^{t}_1,\cdots,\mathbf{x}^{t}_N,\mathbf{c},\boldsymbol{\lambda}^{t}\right),
\\
\mathbf{x}^{t+1}_i&=&\arg \min_{\mathbf{x}_i}\mathcal{L}_\rho\left(\mathbf{x}^{t}_{\neq i},\mathbf{x}_i,\mathbf{c}^{t+1},\boldsymbol{\lambda}^{t}\right)+\frac{1}{2}\left\|\mathbf{x}_i-\mathbf{x}^t_i\right\|^2_{\mathbf{P}_i},\forall i,
\\
\label{eq_lambda_update}
\boldsymbol{\lambda}^{t+1}&=&\boldsymbol{\lambda}^{t}+\beta\rho\left(-\sum_{i=1}^{N}\mathbf{A}_{i}\mathbf{x}^{t+1}_i+\mathbf{b}+\mathbf{c}^{t+1}\right),
\end{IEEEeqnarray}
where $\beta>0$ is a damping parameter, $\mathbf{P}_i$ is a symmetric and positive semi-definite matrix and $\left\|\mathbf{x}_i\right\|^2_{\mathbf{P}_i}\triangleq\mathbf{x}^T_i\mathbf{P}_i\mathbf{x}_i$. Based on the above derivations, the original PM-SLP problem is decomposed, and each subproblem can be calculated in a parallel and distributed manner with (\ref{eq_PJADMM}). The global convergence with $o(1/t)$ convergence rate under certain conditions on $\{\mathbf{P}_i\}$ and $\beta$ of PJ-ADMM can be guaranteed, as established in \cite{deng2017parallel}. In what follows, we derive closed-form solutions for each subproblem in the PJ-ADMM iteration.

\subsection{Closed-Form Solution for Each Subproblem of PJ-ADMM}
\label{subsec_closed-form}
The update for the slack variable $\mathbf{c}$ can be written as
\begin{IEEEeqnarray}{rCl}
\mathbf{c}^{t+1}=\arg \min_{\mathbf{c}\in\mathbb{R}^{2K}_+}\frac{\rho}{2}\left\|-\sum_{i=1}^{N}\mathbf{A}_{i}\mathbf{x}^{t}_i+\mathbf{b}+\mathbf{c}+\frac{\boldsymbol{\lambda}^{t}}{\rho}\right\|^2,
\end{IEEEeqnarray}
which is equivalent to projecting the vector $\sum_{i=1}^{N}\mathbf{A}_{i}\mathbf{x}^{t}_i-\mathbf{b}-\frac{\boldsymbol{\lambda}^{t}}{\rho}$ onto $\mathbb{R}^{2K}_+$, denoted by \begin{IEEEeqnarray}{rCl}P_{\mathbb{R}^{2K}_+}\left(\sum_{i=1}^{N}\mathbf{A}_{i}\mathbf{x}^{t}_i-\mathbf{b}-\frac{\boldsymbol{\lambda}^{t}}{\rho}\right).\nonumber\end{IEEEeqnarray} Its closed-form solution is given by
\begin{IEEEeqnarray}{rCl}
\label{eq_c_update}
\mathbf{c}^{t+1}=\max\left\{\sum_{i=1}^{N}\mathbf{A}_{i}\mathbf{x}^{t}_i-\mathbf{b}-\frac{\boldsymbol{\lambda}^{t}}{\rho},\mathbf{0}\right\}.
\end{IEEEeqnarray}

The iteration for $\mathbf{x}^{t+1}_i$ is updated as follows:
\begin{IEEEeqnarray}{rCl}
\mathbf{x}^{t+1}_i&=&\arg \min_{\mathbf{x}_i} {\left\|\mathbf{x}_i\right\|}^2+\frac{\rho}{2}\left\|-\mathbf{A}_{i}\mathbf{x}_i-\sum_{j\neq i}^{N}\mathbf{A}_{j}\mathbf{x}^{t}_j+\mathbf{b}+\mathbf{c}^{t+1}+\frac{\boldsymbol{\lambda}^{t}}{\rho}\right\|^2+\frac{1}{2}\left\|\mathbf{x}_i-\mathbf{x}^t_i\right\|^2_{\mathbf{P}_i},\forall i,
\IEEEeqnarraynumspace
\end{IEEEeqnarray}
which is an unconstrained quadratic programming, whose optimal solution can be obtained by setting the gradient of the objective function with respect to $\mathbf{x}_i$ to zero, i.e.,
\begin{IEEEeqnarray}{rCl}
2\mathbf{x}_i+\rho\mathbf{A}^T_{i}\left(\mathbf{A}_{i}\mathbf{x}_i+\sum_{j\neq i}^{N}\mathbf{A}_{j}\mathbf{x}^{t}_j-\mathbf{b}-\mathbf{c}^{t+1}-\frac{\boldsymbol{\lambda}^{t}}{\rho}\right)+\mathbf{P}_i\left(\mathbf{x}_i-\mathbf{x}^t_i\right)=0,\forall i.
\end{IEEEeqnarray}
After some calculation, the closed-form solution for $\mathbf{x}^{t+1}_i$ can be written as
\begin{IEEEeqnarray}{rCl}
\label{xiupdate}
\mathbf{x}^{t+1}_i&=&\left(2\mathbf{I}+\rho\mathbf{A}^T_i\mathbf{A}_i+\mathbf{P}_i\right)^{-1}\left[\mathbf{P}_i\mathbf{x}^{t}_i+\rho\mathbf{A}^T_i\left(-\sum_{j\neq i}^{N}\mathbf{A}_{j}\mathbf{x}^{t}_j+\mathbf{b}+\mathbf{c}^{t+1}+\frac{\boldsymbol{\lambda}^{t}}{\rho}\right)\right],\forall i.
\end{IEEEeqnarray}
Note that when we take $N=2N_t$, i.e., the transmit signal vector $\mathbf{x}$ is decomposed into $2N_t$ scalars, $\mathbf{A}_i$ reduces to a column vector $\mathbf{a}_i$, and $\mathbf{P}_i$ reduces to a scalar $p_i$, then the update of the transmit signal can be carried out via $2N_t$ parallel and distributed scalar operations, i.e.,
\begin{IEEEeqnarray}{rCl}
x^{t+1}_i&=&\frac{p_i x^{t}_i+\rho\mathbf{a}^T_i\left(-\sum_{j\neq i}^{2N_t}\mathbf{a}_{j}x^{t}_j+\mathbf{b}+\mathbf{c}^{t+1}+\frac{\boldsymbol{\lambda}^{t}}{\rho}\right)}{2+\rho\mathbf{a}^T_i\mathbf{a}_i+p_i},\forall i.
\end{IEEEeqnarray}
If we group the real and imaginary parts of the same antenna's transmit signal into one block, the transmit signal vector will be decomposed into $N_t$ blocks. Based on the structure of $\mathbf{A}$, we can find that $\mathbf{A}_i\in\mathbb{R}^{2K\times2}$ is a matrix with orthogonal columns, which implies that the corresponding $\mathbf{A}^T_i\mathbf{A}_i$ is a $2\times 2$ diagonal matrix with equal non-zero elements. Therefore, if we take $\mathbf{P}_i$ as a diagonal matrix too, then the matrix inverse operation during the update of $\mathbf{x}_i$ can be replaced by taking the reciprocals of the two entries in the main diagonal with reduced complexity, given by
\begin{IEEEeqnarray}{lCr}
\mathbf{x}^{t+1}_i= \left[\mathbf{P}_i\mathbf{x}^{t}_i+\rho\mathbf{A}^T_i\left(-\sum_{j\neq i}^{N_t}\mathbf{A}_{j}\mathbf{x}^{t}_j+\mathbf{b}+\mathbf{c}^{t+1}+\frac{\boldsymbol{\lambda}^{t}}{\rho}\right)\right]\oslash \mathbf{W},\forall i,
\end{IEEEeqnarray}
where $\mathbf{W}\triangleq diag\left(2\mathbf{I}+\rho\mathbf{A}^T_i\mathbf{A}_i+\mathbf{P}_i\right)$.

\subsection{Convergence Analysis}
\label{convergenceAnalysis}
The global convergence of the PJ-ADMM for linear equality constraints is established in \cite{deng2017parallel}. As shown in the preceding section, the linear inequality constraints of PM-SLP are reformulated into linear equality constraints with the aid of the slack variable $\mathbf{c}$, which is an exactly updated intermediate variable, thus not affecting convergence \cite{gabay1976dual}. For the sake of completeness, the global convergence theorem of the PJ-ADMM PM-SLP is stated in the following.
\newtheorem{theorem}{\bf Theorem}
\begin{theorem}
Let $\{\mathbf{u}^t\}$ be the sequence generated by (\ref{eq_PJADMM}) with arbitrary initialization. If there exists $\epsilon_i>0$ such that
\begin{IEEEeqnarray}{rCl}
\mathbf{P}_i&\succeq&\rho(\frac{1}{\epsilon_i}-1)\mathbf{A}^T_i\mathbf{A}_i, \forall i, \sum^N_{i=1}\epsilon_i \leq 2-\beta,
\end{IEEEeqnarray}
then $\{\mathbf{u}^t\}$ converges to a solution $\mathbf{u}^*$ to the PM-SLP problem.
\end{theorem}
\begin{IEEEproof}[\bf Proof]
See Appendix \ref{proof}.
\end{IEEEproof}

Furthermore, by choosing $\epsilon_i\leq\frac{2-\beta}{N}$, the sufficient condition can be rewritten as
\begin{IEEEeqnarray}{rCl}
\label{sufficient condition}
\mathbf{P}_i\succeq\rho(\frac{N}{2-\beta}-1)\mathbf{A}^T_i\mathbf{A}_i, \forall i.
\end{IEEEeqnarray}
There are two special choices for $\mathbf{P}_i$ as mentioned in \cite{deng2017parallel}. The first one is termed the standard proximal, which takes the following form:
\begin{IEEEeqnarray}{rCl}
\label{standard proximal}
\mathbf{P}_i=\tau_i\mathbf{I}, 
\end{IEEEeqnarray}
where $\tau_i\geq\rho\left(\frac{N}{2-\beta}-1\right)\left\|\mathbf{A}_i\right\|^2$. The other is termed the prox-linear proximal, which takes the following form:
\begin{IEEEeqnarray}{rCl}
\label{prox-linear proximal}
\mathbf{P}_i=\tau_i\mathbf{I}-\rho\mathbf{A}^T_i\mathbf{A}_i, 
\end{IEEEeqnarray}
where $\tau_i\geq\frac{\rho N}{2-\beta}\left\|\mathbf{A}_i\right\|^2$.

\subsection{Adaptive Parameter Tuning Strategy}
\label{adaptiveStrategy}
In the previous section, we have derived a sufficient condition to guarantee the convergence of the proposed algorithm, which provides a lower bound for the Hessian of the proximal term $\mathbf{P}_i$. However, the basic inequality (\ref{basic inequality}) for bounding $\left\|\mathbf{u}\right\|^2_{\mathbf{Q}}$ is usually rather loose, so the sufficient condition may be fairly conservative in practical implementation \cite{deng2017parallel}.

In order to accelerate convergence, compared to adopting a constant proximal coefficient that satisfies the sufficient condition (\ref{sufficient condition}), it is more preferred to initialize $\mathbf{P}_i$ with a relatively small value and increase it iteratively until $\|\mathbf{u}^t-\mathbf{u}^{t+1}\|^2_{\mathbf{Q}}\geq 0$, i.e., adaptive tuning the proximal coefficient matrix $\mathbf{P}_i$. \cite{deng2017parallel} proposed a heuristic scheme to tune the proximal coefficient matrix $\mathbf{P}_i$ based on the exact value of $\|\mathbf{u}^t-\mathbf{u}^{t+1}\|^2_{\mathbf{Q}}$, specifically, the proximal coefficient is increased when $\mathbf{Q}$ is not positive semi-definite, otherwise it will remain constant. The aforementioned analysis indicates that the resulting constant proximal coefficient does not necessarily satisfy the sufficient condition (\ref{sufficient condition}). The details of the adaptive parameter tuning scheme for our proposed PJ-ADMM are summarized as follows:
\begin{IEEEeqnarray}{rCl}
\mathbf{P}^{t+1}_i=
\begin{cases}
\delta_i\mathbf{P}^{t}_i,&\text{if } \|\mathbf{u}^{t-1}-\mathbf{u}^t\|^2_{\mathbf{Q}}<\eta\|\mathbf{u}^{t-1}-\mathbf{u}^t\|^2,\\
\mathbf{P}^{t}_i, &\text{otherwise,}
\end{cases}
\IEEEeqnarraynumspace
\end{IEEEeqnarray}
where $\delta_i>1$, $\eta>0$ is a sufficient small scalar.
\subsection{Efficient and Inverse-Free Algorithms}
\label{efficientAlgr}
For the standard proximal term (\ref{standard proximal}), when $\mathbf{A}^T_i\mathbf{A}_i$ is a nondiagonal matrix, and the proximal parameter $\tau_i$ is tuned within each iteration, matrix inverse operation is required whenever $\tau_i$ is changed, thus the computational complexity of the algorithm is dominated by matrix inversion. The overall computational complexity can be further reduced by circumventing matrix inverse operation, which can be realized based on the fact that the eigenspace of the matrix to be inverted $\rho\mathbf{A}^T_i\mathbf{A}_i+(2+\tau_i)\mathbf{I}$ is inherited from the eigenspace of $\mathbf{A}^T_i\mathbf{A}_i$ \cite{chu2019efficient}. To be more specific, we firstly express the singular value decomposition (SVD) of $\mathbf{A}^T_i\mathbf{A}_i$ as
\begin{IEEEeqnarray}{rCl}
\mathbf{A}^T_i\mathbf{A}_i=\mathbf{U}_i\mathbf{\Sigma}_i\mathbf{V}^T_i,
\end{IEEEeqnarray}
where $\mathbf{U}_i$, $\mathbf{V}_i$ are the right and left singular matrix, respectively, and $\mathbf{U}_i$ is a unity matrix, i.e., $\mathbf{U}^{-1}_i=\mathbf{U}^T_i$. From the symmetry, we have $\mathbf{V}_i=\mathbf{U}_i$. $\mathbf{\Sigma}_i$ is a diagonal matrix of which the diagonal entries are singular values of $\mathbf{A}^T_i\mathbf{A}_i$. Based on the above, each $\mathbf{x}_i$ can be efficiently updated by
\begin{IEEEeqnarray}{rCl}
\mathbf{x}^{t+1}_i&=&\left(2\mathbf{I}+\rho\mathbf{A}^T_i\mathbf{A}_i+\tau_i\mathbf{I} \right)^{-1}\mathbf{r}_i=\mathbf{U}_i\left[\rho\mathbf{\Sigma}_i+(2+\tau_i)\mathbf{I}\right]^{-1}\mathbf{U}^T_i\mathbf{r}_i=\mathbf{U}_i \left(\mathbf{U}^T_i \mathbf{r}_i \oslash\mathbf{q}_i\right), \forall i,\IEEEeqnarraynumspace
\end{IEEEeqnarray}
where
$
\mathbf{q}_i\triangleq  diag \{\rho\mathbf{\Sigma}_i+(2+\tau_i)\mathbf{I}\},
\mathbf{r}_i\triangleq \tau_i\mathbf{x}^{t}_i+\rho\mathbf{A}^T_i \left(-\sum_{j\neq i}^{N}\mathbf{A}_{j}\mathbf{x}^{t}_j+\mathbf{b}+\mathbf{c}^{t+1}+\frac{\boldsymbol{\lambda}^{t}}{\rho}\right).
$

With the above approach, the matrix inversion in each update of $\mathbf{x}_i$ is replaced by one SVD in the first update and incremental matrix-vector multiplications in the remaining updates.

So far, the computational complexity induced by the adaptive parameter tuning strategy is alleviated by the preceding SVD-based efficient algorithm, thereby the constant proximal and the adaptive proximal only need one matrix inverse operation or SVD, respectively, both with $\mathcal{O}((2N_t)^3)$ complexity. For a massive MU-MIMO system equipped with hundreds of downlink transmit antennas, such complexity reduction is remarkable.

As a step further, we can observe that the matrix inversion is needed by the non-diagonal coefficient matrix $\rho\mathbf{A}^T_i\mathbf{A}_i$ of the quadratic penalty term in the augmented Lagrangian function (\ref{augmented Lagrangian}). Fortunately, flexible as the proximal term is, it can be used to subtract $\rho\mathbf{A}^T_i\mathbf{A}_i$, which means the matrix inverse operation can be fully eliminated. To devise such an inverse-free algorithm, we propose to construct a prox-linear proximal term as in (\ref{prox-linear proximal}), which linearizes the quadratic penalty term by approximating the Hessian $\rho\mathbf{A}^T_i\mathbf{A}_i$ of the quadratic penalty term with an identity proximal matrix $\tau_i \mathbf{I}$. Accordingly, the inverse-free closed-form solutions for the update of $\mathbf{x}_i$ can be obtained by substituting (\ref{prox-linear proximal}) into (\ref{xiupdate}), i.e.,
\begin{IEEEeqnarray}{rCl}
\mathbf{x}^{t+1}_i&=&\frac{1}{2+\tau_i}
\left[\tau_i\mathbf{x}^{t}_i+\rho\mathbf{A}^T_i\left(-\mathbf{A}\mathbf{x}^{t}+\mathbf{b}+\mathbf{c}^{t+1}+\frac{\boldsymbol{\lambda}^{t}}{\rho}\right)\right],\forall i.
\end{IEEEeqnarray}
The computational complexity is mainly induced by the matrix-vector multiplications, any single matrix inversion or SVD is no longer demanded.

\subsection{Decentralized Algorithm for Low-Coordination Overhead}
\begin{figure}[!t]
\centering
\subfloat[Proposed parallel SLP approach with one centralized consensus node.]{
\includegraphics{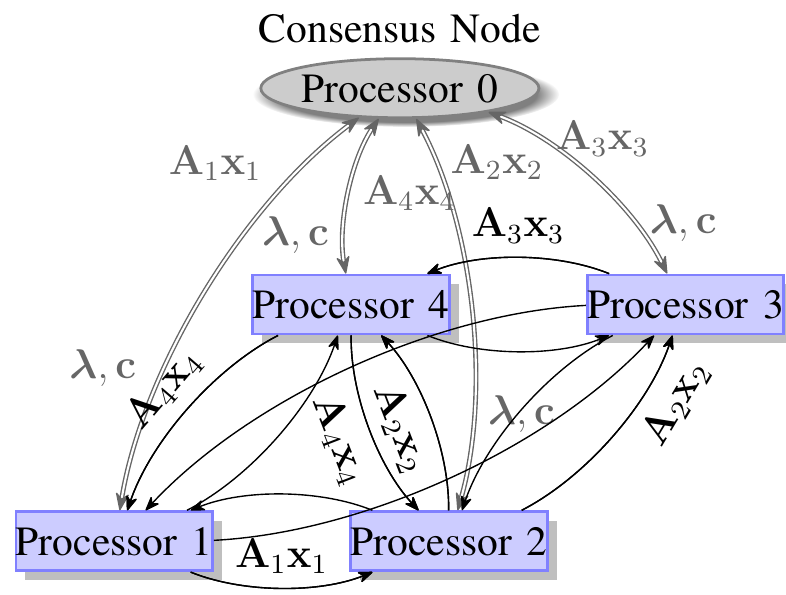}
\label{fig_centralized}
}
\hfil
\subfloat[Proposed parallel SLP approach with fully decentralized nodes.]{
\includegraphics{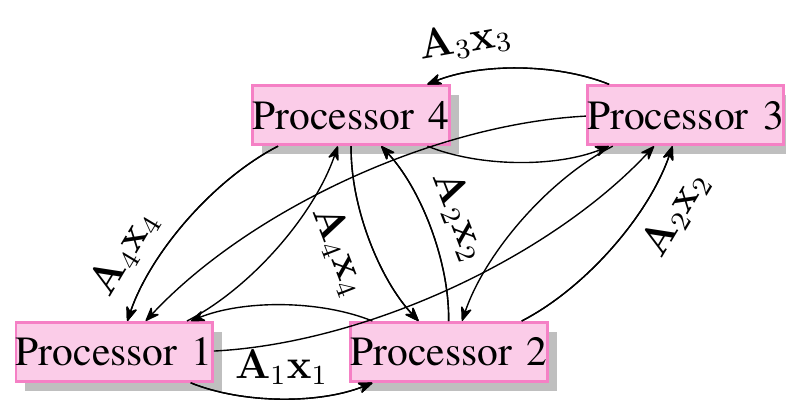}
\label{fig_decentralized}
}
\caption{An example of parallel centralized and decentralized SLP systems.}
\label{fig_parallelBlock}
\end{figure}

We can envision that the PIF-SLP algorithm can be implemented by a network of processor units connected by communication links \cite{bertsekas2015parallel}, which will consume extra resources and cause time delay. In the preceding PJ-ADMM algorithm, iterations for the slack variable and Lagrangian multiplier vector need to be carried out at a central node, which requires extra information exchange with other processor units in the system to aggregate and propagate intermediate results. In such case, the time spent in exchanging information cannot be neglected. When real-time implementation and low-latency communication is required, we can reduce the coordination overhead by incorporating the slack variable iteration into the primal and dual variable iterations, and carrying out the multiplier iteration at the $N$ blocks of transmit signal vector iterations. Another valuable feature is that the central node is no longer required, enabling us to further come up with a low-coordination overhead decentralized PIF-SLP algorithm. Without loss of generality, an example illustration of the centralized, as well as the decentralized system with $N=4$ parallel processing units is shown in Fig. \ref{fig_parallelBlock}. For the centralized scheme shown in Fig. \ref{fig_centralized}, the processor unit 0 is working as a consensus node that collects the parallel processor units' results $\{\mathbf{A}_i\mathbf{x}_i\}$. As shown in Fig. \ref{fig_decentralized}, consisting of the underlying 4 fully parallel processor units, the decentralized scheme does not need a central node and enjoys the lower coordination overhead.

For notation simplicity, we first denote $g\left(\mathbf{x}\right)\triangleq-\sum^{N}_{i=1}\mathbf{A}_{i}\mathbf{x}_i+\mathbf{b}$. The closed-form solution (\ref{eq_c_update}) for the slack variable $\mathbf{c}$ can be rewritten as
\begin{IEEEeqnarray}{rCl}
\mathbf{c}^{t+1}=\max\left\{-g\left(\mathbf{x}^{t}\right)-\frac{\boldsymbol{\lambda}^{t}}{\rho},\mathbf{0}\right\}.
\end{IEEEeqnarray}
Denoting $g^+\left(\mathbf{x},\boldsymbol{\lambda},\rho\right)\triangleq\max\left\{g\left(\mathbf{x}\right),-\frac{\boldsymbol{\lambda}}{\rho}\right\}$, we have 
\begin{IEEEeqnarray}{rCl}
\label{g+equality}
g^+\left(\mathbf{x},\boldsymbol{\lambda},\rho\right)=g\left(\mathbf{x}\right)+\mathbf{c}.
\end{IEEEeqnarray}
Substituting (\ref{g+equality}) into (\ref{augmented Lagrangian}), we can rewrite the augmented Lagrangian function as 
\begin{IEEEeqnarray}{rCl}
\mathcal{L}_\rho\left(\mathbf{x},\boldsymbol{\lambda}\right)&=&\sum_{i=1}^{N}\left\|\mathbf{x}_i\right\|^2+\boldsymbol{\lambda}^T\mathbf{g}^+\left(\mathbf{x},\boldsymbol{\lambda},\rho\right)+\frac{\rho}{2}\left\|\mathbf{g}^+\left(\mathbf{x},\boldsymbol{\lambda},\rho\right)\right\|^2
\nonumber\\
&=&\sum_{i=1}^{N}\left\|\mathbf{x}_i\right\|^2+\frac{1}{2\rho}\left(\left\|\max\left\{\boldsymbol{\lambda}+\rho g\left(\mathbf{x}\right),\mathbf{0}\right\}\right\|^2-\left\|\boldsymbol{\lambda}\right\|^2\right),
\end{IEEEeqnarray}
where the penalty term $\frac{1}{2\rho}\left(\left\|\max\left\{\boldsymbol{\lambda}+\rho g\left(\mathbf{x}\right),\mathbf{0}\right\}\right\|^2-\left\|\boldsymbol{\lambda}\right\|^2\right)$ corresponding to the inequality CI constraints is continuously differentiable with respect to $\mathbf{x}$ as $g\left(\mathbf{x}\right)$ is continuously differentiable \cite{bertsekas2016nonlinear}. Following the preceding procedure, we can obtain closed-form expressions similar to the preceding PJ-ADMM. By substituting (\ref{g+equality}) into the update for $\mathbf{x}_i$ in (\ref{xiupdate}) and $\boldsymbol{\lambda}$ in (\ref{eq_lambda_update}), $\mathbf{x}^{t+1}_i$ and $\boldsymbol{\lambda}^{t+1}$ can be further expressed as
\begin{IEEEeqnarray}{rCl}
\label{communication-efficient}
\IEEEyesnumber\IEEEyessubnumber*
\mathbf{x}^{t+1}_i&=&\left(2\mathbf{I}+\rho\mathbf{A}^T_i\mathbf{A}_i+\mathbf{P}_i\right)^{-1}\left[\mathbf{P}_i\mathbf{x}^{t}_i+\rho\mathbf{A}^T_i\left(\mathbf{g}^+\left(\mathbf{x}^{t},\boldsymbol{\lambda}^t,\rho\right)+\mathbf{A}_{i}\mathbf{x}^{t}_i+\frac{\boldsymbol{\lambda}^{t}}{\rho}\right)\right],\forall i,
\\
\boldsymbol{\lambda}^{t+1}&=&\boldsymbol{\lambda}^{t}+\beta\rho\mathbf{g}^+\left(\mathbf{x}^{t+1},\boldsymbol{\lambda}^t,\rho\right).
\end{IEEEeqnarray}
A slight difference between PJ-ADMM and its decentralized counterpart during iteration lies in that the latter first uses the updated transmit signal vector to reformulate the slack variable $\mathbf{c}$, based on which the multiplier vector $\boldsymbol{\lambda}$ is updated.

For completeness, we further propose the decentralized PIF-SLP algorithm, which can be obtained by substituting the prox-linear proximal term (\ref{prox-linear proximal}) into (\ref{communication-efficient}), given by
\begin{IEEEeqnarray}{rCl}
\IEEEyesnumber\IEEEyessubnumber*
\label{CEIFx}
\mathbf{x}^{t+1}_i&=&\frac{1}{2+\tau_i}\left(\tau_i\mathbf{x}^{t}_i+\rho\mathbf{A}^T_i\left(\mathbf{g}^+\left(\mathbf{x}^{t},\boldsymbol{\lambda}^t,\rho\right)+\frac{\boldsymbol{\lambda}^{t}}{\rho}\right)\right),\forall i,
\\
\label{CEIFlambda}
\boldsymbol{\lambda}^{t+1}&=&\boldsymbol{\lambda}^{t}+\beta\rho\mathbf{g}^+\left(\mathbf{x}^{t+1},\boldsymbol{\lambda}^t,\rho\right).
\end{IEEEeqnarray}
The corresponding algorithm is summarized in Algorithm \ref{alg:PIF-SLP}.
\begin{figure}[!t]
		\renewcommand{\algorithmicrequire}{\textbf{Input:}}
		\renewcommand{\algorithmicensure}{\textbf{Output:}}
		\removelatexerror
		\begin{algorithm}[H]
			\caption{Proposed Low-Coordination Overhead Decentralized PIF-SLP Algorithm}
			\label{alg:PIF-SLP}
			\begin{algorithmic}[1]
				\REQUIRE $\mathbf{A}$, $\mathbf{b}$, $\rho$, $\eta$, $\{\delta_i\}^N_{i=1}$          
				\ENSURE $\mathbf{x}$  
				\STATE Initialize $\mathbf{x}^0_i\;(i=1,\cdots,N), \boldsymbol{\lambda}^0$ and $\tau^0_i\;(i=1,\cdots,N)$;
				\FOR {$t=0,1,\cdots$}		
				\STATE Update $\mathbf{x}^{t+1}_i$ for $i=1,\cdots,N$ in parallel by (\ref{CEIFx});
				
				\STATE Share $\mathbf{A}_i\mathbf{x}^{t+1}_i$;
				\STATE Collect $\{\mathbf{A}_j\mathbf{x}^{t+1}_j\}_{j\neq i}$;
				\STATE Update $\boldsymbol{\lambda}^{t+1}$ for $i=1,\cdots,N$ in parallel by (\ref{CEIFlambda});
				\IF {$\|\mathbf{u}^{t}-\mathbf{u}^{t+1}\|^2_{\mathbf{Q}}<\eta\|\mathbf{u}^{t}-\mathbf{u}^{t+1}\|^2$}
				\STATE $\tau_i\leftarrow\delta_i\tau_i$;
				\STATE Backtrack $\mathbf{u}^{t+1}\leftarrow\mathbf{u}^{t}$; 
				\ENDIF
				\ENDFOR
			\end{algorithmic}
		\end{algorithm}
	\end{figure}

\section{Complexity and Overhead Analysis}
\label{complexity}
\subsection{Computational Complexity}
We evaluate the computational complexity of the proposed parallel SLP approach, including the matrix inversion-based plain implementation, the SVD-based efficient algorithm, and the PIF-SLP algorithm by counting float count operations in this section. Define the flop-count operator $\mathcal{F}\left(\mathbf{z}|\mathbf{y}\right)$ as the number of flops to compute $\mathbf{z}$ given $\mathbf{y}$.
Assume that the dimensions of the parallel subproblems are equal, and let $m,n$ be the row and column dimensions of $\mathbf{A}_i$ respectively, i.e., $m=2K,n=2N_t/N$.
When no matrix structure is exploited, matrix inversion is required in each iteration, then the straight and naive implementation of the proposed parallel PM-SLP approach costs
\begin{IEEEeqnarray}{rCl}
\mathcal{F}\left(\boldsymbol{\lambda}^{t+1}|\boldsymbol{\lambda}^t\right)&=&\mathcal{F}\left(\mathbf{c}^{t+1}|\left(\mathbf{A}_i\mathbf{x}^{t}_i,\boldsymbol{\lambda}^{t}\right)\right)+\mathcal{F}\left(\mathbf{A}_i\mathbf{x}^{t+1}_i|\left(\mathbf{A}_i\mathbf{x}^{t}_i,\mathbf{c}^{t+1},\boldsymbol{\lambda}^t\right)\right)\nonumber\\&&+\mathcal{F}\left(\boldsymbol{\lambda}^{t+1}|\left(\boldsymbol{\lambda}^t,\mathbf{A}_i\mathbf{x}^{t+1}_i,\mathbf{c}^{t+1}\right)\right)\nonumber\\
&=&\mathcal{O}(m)+\mathcal{O}((m+n)n^2)+\mathcal{O}((m+n)n)+\mathcal{O}(m)
\end{IEEEeqnarray}
flops per iteration. The dominant terms are caused by matrix inversion, matrix-matrix multiplication, and matrix-vector multiplication during the update of $\mathbf{x}_i$. As discussed in Section \ref{efficientAlgr}, when SVD is precomputed and used in the subsequent iterations, then the SVD-based efficient algorithm costs
\begin{IEEEeqnarray}{rCl}
\mathcal{F}\left(\boldsymbol{\lambda}^{t+1}|\boldsymbol{\lambda}^t\right)&=&\mathcal{F}\left(\mathbf{c}^{t+1}|\left(\mathbf{A}_i\mathbf{x}^{t}_i,\boldsymbol{\lambda}^{t}\right)\right)+\mathcal{F}\left(\mathbf{A}_i\mathbf{x}^{t+1}_i|\left(\mathbf{A}_i\mathbf{x}^{t}_i,\mathbf{c}^{t+1},\boldsymbol{\lambda}^t,\mathbf{U}_i,\mathbf{\Sigma}_i\right)\right)\nonumber\\&&+\mathcal{F}\left(\boldsymbol{\lambda}^{t+1}|\left(\boldsymbol{\lambda}^t,\mathbf{A}_i\mathbf{x}^{t+1}_i,\mathbf{c}^{t+1}\right)\right)\nonumber\\
&=&\mathcal{O}(m)+\mathcal{O}((m+n)n)+\mathcal{O}(m)
\end{IEEEeqnarray}
flops per iteration. The dominant terms of the efficient implementation turn to matrix-vector multiplication since matrix inversion and matrix-matrix multiplication are both eliminated. The PIF-SLP algorithm costs
\begin{IEEEeqnarray}{rCl}
\mathcal{F}\left(\boldsymbol{\lambda}^{t+1}|\boldsymbol{\lambda}^t\right)&=&\mathcal{F}\left(\mathbf{c}^{t+1}|\left(\mathbf{A}_i\mathbf{x}^{t}_i,\boldsymbol{\lambda}^{t}\right)\right)+\mathcal{F}\left(\mathbf{A}_i\mathbf{x}^{t+1}_i|\left(\mathbf{A}_i\mathbf{x}^{t}_i,\mathbf{c}^{t+1},\boldsymbol{\lambda}^t\right)\right)\nonumber\\&&+\mathcal{F}\left(\boldsymbol{\lambda}^{t+1}|\left(\boldsymbol{\lambda}^t,\mathbf{A}_i\mathbf{x}^{t+1}_i,\mathbf{c}^{t+1}\right)\right)\nonumber\\
&=&\mathcal{O}(m)+\mathcal{O}((m+1)n)+\mathcal{O}(m)
\end{IEEEeqnarray}
flops per iteration. It can be observed that the PIF-SLP algorithm is not only free of matrix inversion or SVD but also requires fewer matrix multiplications compared to the other two algorithms.

\subsection{Coordination Overhead}
For the parallel SLP scheme with a consensus node, the iteration needs $N+1$ processor units, of which $N$ for the update of $\mathbf{x}_i$, and the extra one for multiplier and slack variable update. Assume the CSI of transmit antennas is only accessed by the corresponding processors, while the data information is known by all processor units. The algorithm requires sharing CSI and interim results among processor units. We share $\{\mathbf{A}_i\mathbf{x}^t_i\}$ instead of sharing $\{\mathbf{A}_i\}$ and $\{\mathbf{x}^t_i\}$ separately, for reduced coordination overhead. Hence, the coordination overhead per iteration of the $N+1$ processor units is $\mathcal{Q}N(N+2)m$ bits, where $\mathcal{Q}$ denotes the required bits for exchanging one real-valued scalar.

As for the low-coordination overhead decentralized counterpart, the processor dedicated to consensus variables, i.e., multiplier and slack variable, is eliminated. The multiplier is updated in each local processor unit, therefore the need for the exchange of consensus variables is removed. The exchanged information is the sole $
\{\mathbf{A}_i\mathbf{x}^t_i\}$, thus the coordination overhead per iteration of the $N$ processor units is $\mathcal{Q}N(N-1)m$ bits.

\section{Numerical Results}
\label{results}
This section evaluates and compares the performance of the proposed algorithms via Monte Carlo simulations. We assume each user has unit noise variance and equal instantaneous SINR threshold, i.e., $\sigma^2_k=\sigma^2=1$, $\gamma_k=\gamma,\forall k$. QPSK modulation is employed throughout the simulations. A downlink massive MU-MISO system with 128 transmit antennas to serve 112 single-antenna users is considered unless otherwise specified. The transmit signal vector is partitioned into 64 blocks, namely $N=64$, with 4 elements in each block.

In the sequel for clarity, we list the proposed algorithms as well as the benchmark schemes we have compared in our simulations:
\begin{enumerate}[\IEEEsetlabelwidth{12)}]
\item ‘ZF': The conventional ZF scheme with symbol-level power normalization. The corresponding precoded signal vector is given by
\begin{IEEEeqnarray}{rCl}
\tilde{\mathbf{x}}_{ZF}=\frac{1}{f_{ZF}}\tilde{\mathbf{H}}^H\left(\tilde{\mathbf{H}}\tilde{\mathbf{H}}^H\right)^{-1}\tilde{\mathbf{s}},
\end{IEEEeqnarray}
where $f_{ZF}$ is the symbol-level scaling factor. For the sake of comparison, the ZF transmit signal is normalized by the transmit power obtained by the IPM, thus we have
\begin{IEEEeqnarray}{rCl}
f_{ZF}=\frac{\left\|\tilde{\mathbf{H}}^H\left(\tilde{\mathbf{H}}\tilde{\mathbf{H}}^H\right)^{-1}\tilde{\mathbf{s}}\right\|}{\left\|\tilde{\mathbf{x}}_{IPM}\right\|},
\end{IEEEeqnarray}
where $\tilde{\mathbf{x}}_{IPM}$ is the complex-valued precoded signal obtained by the IPM for PM-SLP.
\item ‘RZF': The conventional RZF scheme with symbol-level power normalization. The corresponding precoded signal vector is given by
\begin{IEEEeqnarray}{rCl}
\tilde{\mathbf{x}}_{RZF}=\frac{1}{f_{RZF}}\tilde{\mathbf{H}}^H\left(\tilde{\mathbf{H}}\tilde{\mathbf{H}}^H+\varphi\mathbf{I}\right)^{-1}\tilde{\mathbf{s}},
\end{IEEEeqnarray}
where $f_{RZF}$ is the symbol-level scaling factor, $\varphi$ is the regulation parameter, which is set to $\varphi=\sigma^2$. The RZF transmit signal is also normalized by the transmit power obtained by the IPM, thus we have
\begin{IEEEeqnarray}{rCl}
f_{RZF}=\frac{\left\|\tilde{\mathbf{H}}^H\left(\tilde{\mathbf{H}}\tilde{\mathbf{H}}^H+\varphi\mathbf{I}\right)^{-1}\tilde{\mathbf{s}}\right\|}{\left\|\tilde{\mathbf{x}}_{IPM}\right\|}.
\end{IEEEeqnarray}
\item ‘IPM': The IPM for PM-SLP \cite{grant2014cvx}.
\item ‘EGPA': The efficient gradient projection algorithm for PM-SLP \cite{masouros2015exploiting}.
\item ‘SCF': The suboptimal closed-form solution for PM-SLP \cite{haqiqatnejad2018power}.
\item ‘ISCF': The improved suboptimal closed-form solution for PM-SLP \cite{haqiqatnejad2019approximate}.
\item ‘PSLP-SA': The proposed parallel and distributed approach for PM-SLP, with the standard proximal term and the adaptive parameter tuning strategy.
\item ‘PSLP-SC': The proposed parallel and distributed approach for PM-SLP, with the standard proximal term and constant parameters.
\item ‘PSLP-LA': The proposed parallel and distributed approach for PM-SLP, with the prox-linear proximal term and the adaptive parameter tuning strategy.
\item ‘PSLP-LC': The proposed parallel and distributed approach for PM-SLP, with the prox-linear proximal term and constant parameters.
\end{enumerate}

For PSLP-SA and PSLP-LA, the proximal parameters are initialized as $\tau_i=0.1(N-1)\rho$ and adaptively updated by the adaptive parameter tuning strategy in Section \ref{adaptiveStrategy} with $\delta_i=2$. For PSLP-SC and PSLP-LC, we choose $\tau_i=0.2\rho\left(\frac{N}{2-\beta}-1\right)\left\|\mathbf{A}_i\right\|^2$. The penalty parameter $\rho$ is set to be 0.06; the damping parameter $\beta$ is set to be 1.

\subsection{Convergence Behavior}
\begin{figure}[!t]
\centering
\subfloat[Iterate gap]{\includegraphics[width=18pc]{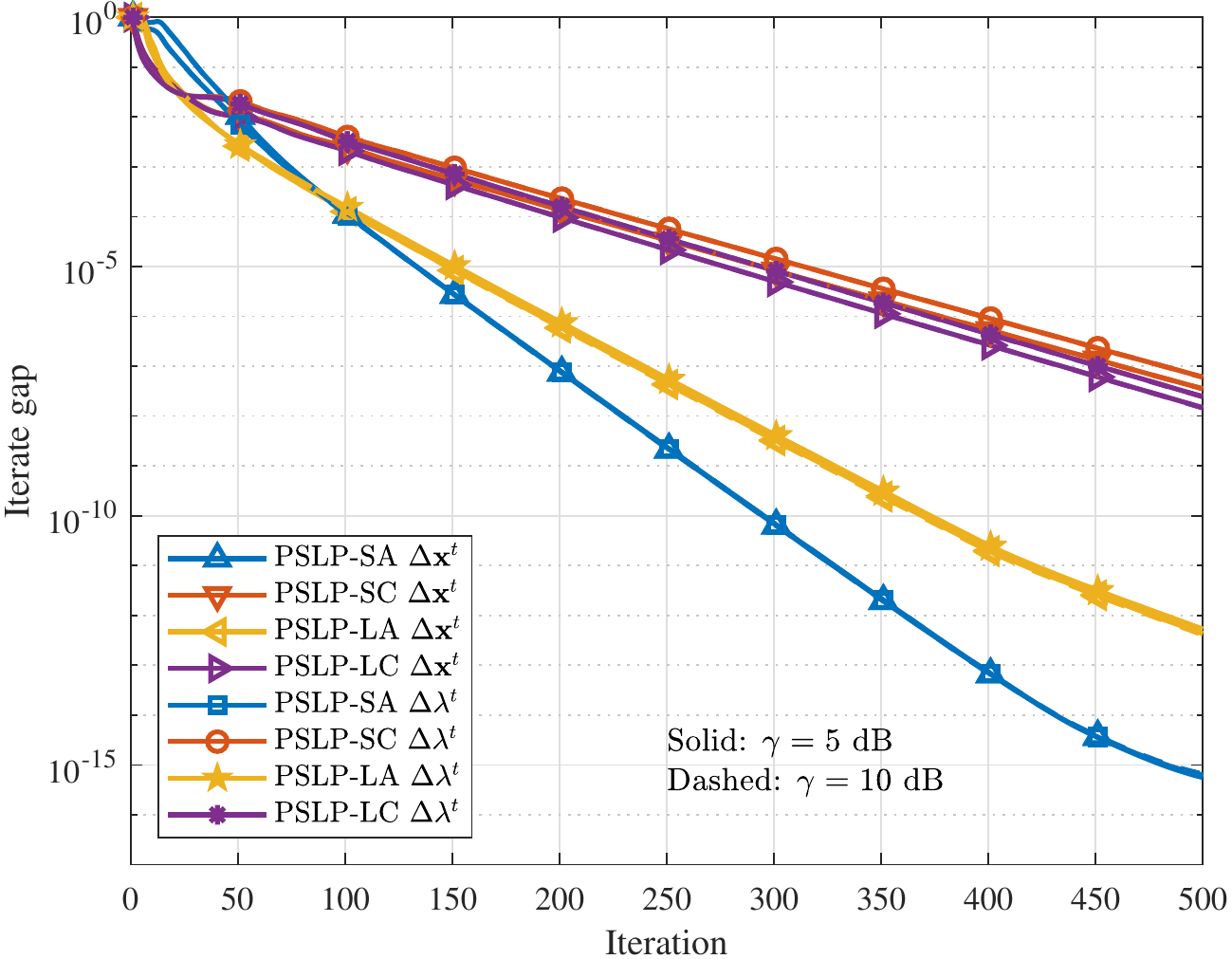}
\label{fig_first_case_itergap}}
\hfil
\subfloat[Average transmit power]{\includegraphics[width=18pc]{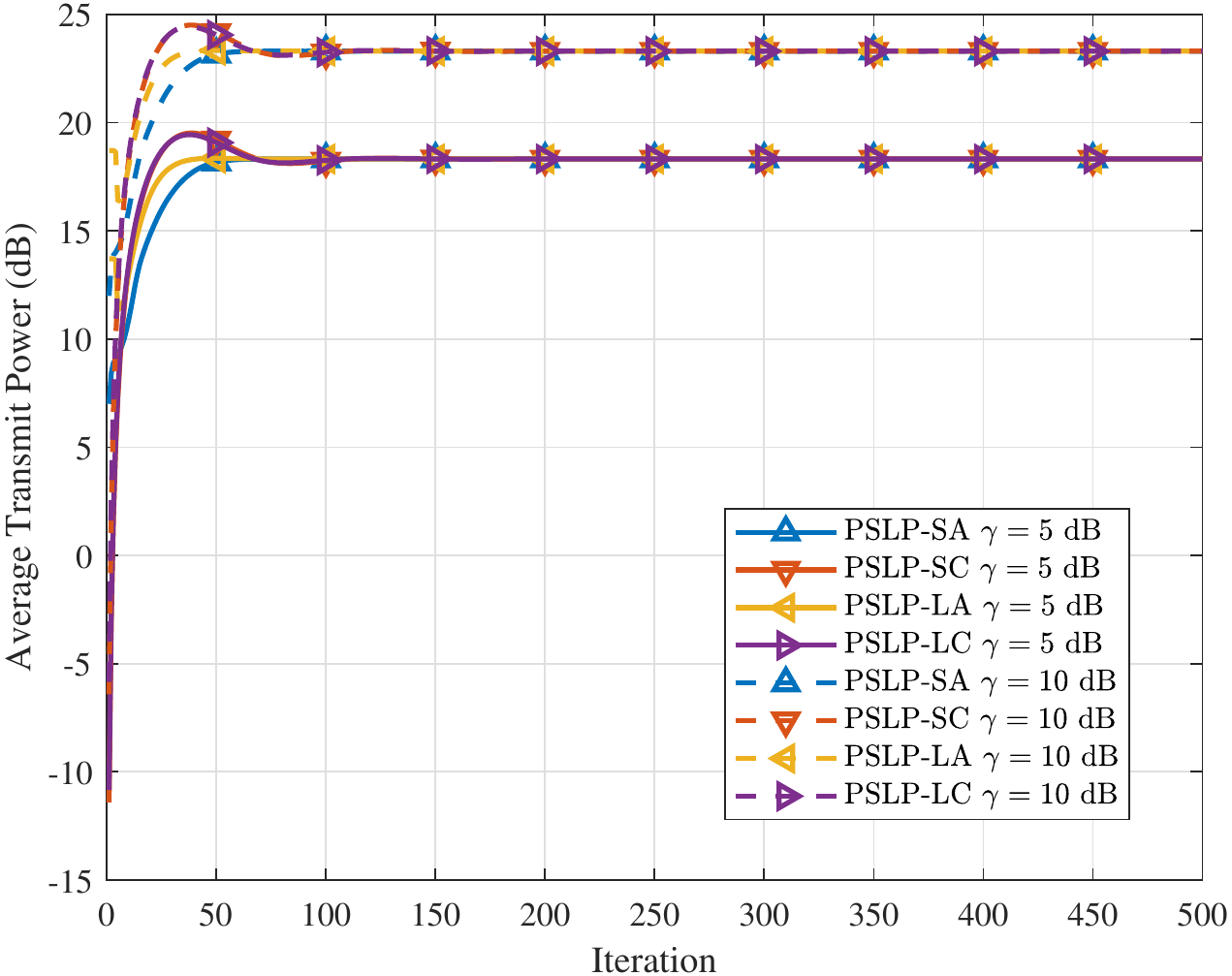}
\label{fig_second_case_convPower}}
\hfil
\caption{Convergence behavior of the proposed approach in different SINR thresholds, QPSK, $N_t=128$, $K=112$, $N=64$.}
\label{fig_converge}
\end{figure}
We first demonstrate the convergence behavior of the proposed approach. As the adaptive parameter tuning strategy needs precise information to evaluate whether the parameters need to be tuned, we adopt the more accurate low-coordination overhead decentralized formulation. The iterate gap of the primal and dual variable is defined as
$
\IEEEyesnumber\IEEEyessubnumber*
\Delta\mathbf{x}^t\triangleq \left\|\mathbf{x}^t-\mathbf{x}^{t-1}\right\| /\left\|\mathbf{x}^t\right\|,
\Delta\boldsymbol{\lambda}^t\triangleq\left\|\boldsymbol{\lambda}^t-\boldsymbol{\lambda}^{t-1}\right\| /\left\|\boldsymbol{\lambda}^t\right\|.
$

Fig. \ref{fig_converge} illustrates the convergence behavior of the proposed algorithms in terms of iterate gap and average transmit power, which are both averaged over 2000 random channel realizations. The results in Fig. \ref{fig_converge} show that the sequence generated by the proposed algorithm is convergent to a unique solution. The alternatively optimized primal and dual variable has an identical iterate gap. The algorithms with adaptive parameter tuning have a faster convergence rate compared to the algorithms with constant parameters. Meanwhile, using the same parameters, the inverse-free algorithms with the prox-linear proximal term have slightly slower convergence performance than the more complex algorithms with the standard proximal term.

\subsection{Transmit Power and Uncoded BER Performance}
\begin{figure}[!t]
\centering
\subfloat[Average transmit power v.s. SINR threshold]{\includegraphics[width=18pc]{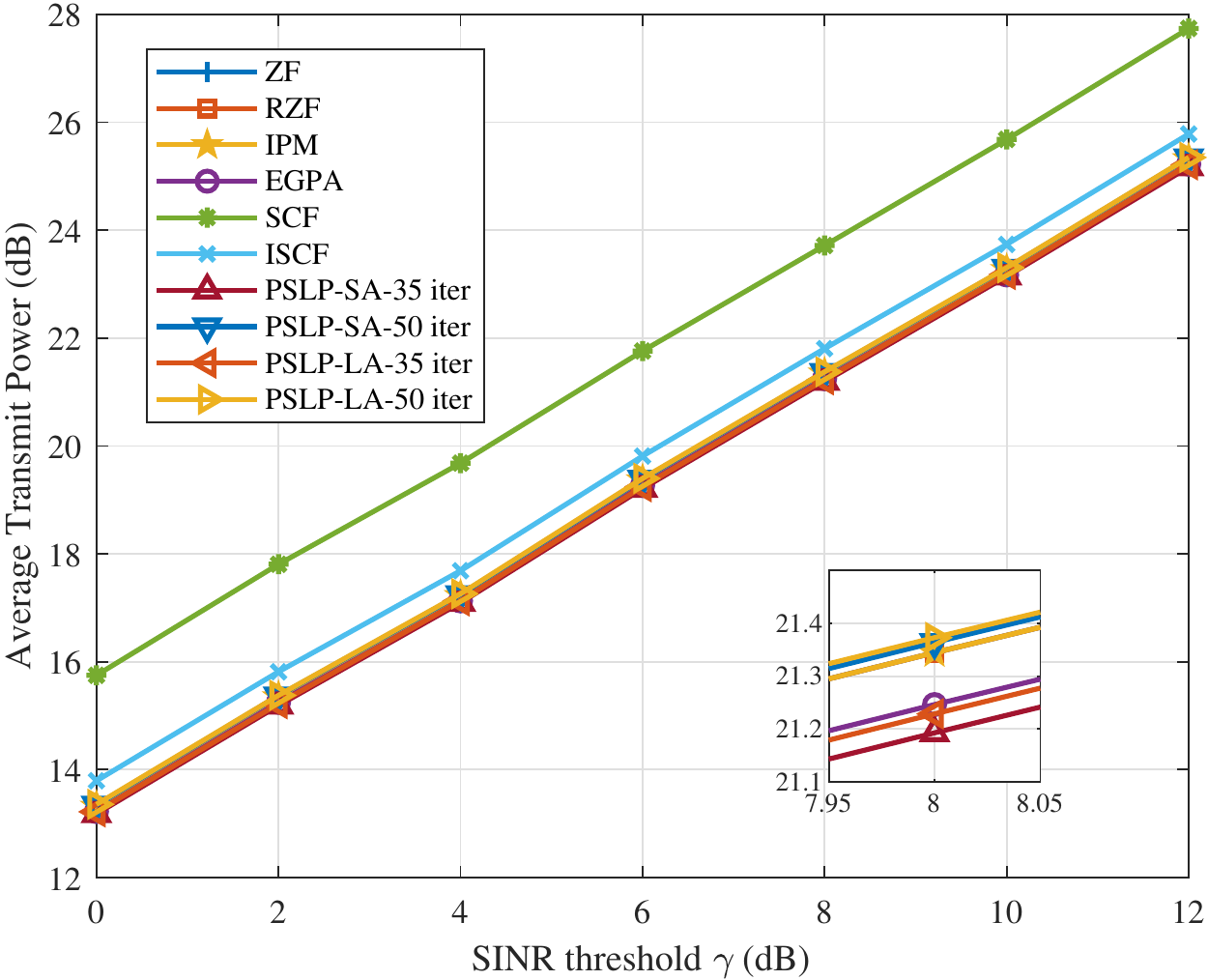}
\label{fig_first_case_power}}
\hfil
\subfloat[Uncoded BER v.s. SINR threshold]{\includegraphics[width=18pc]{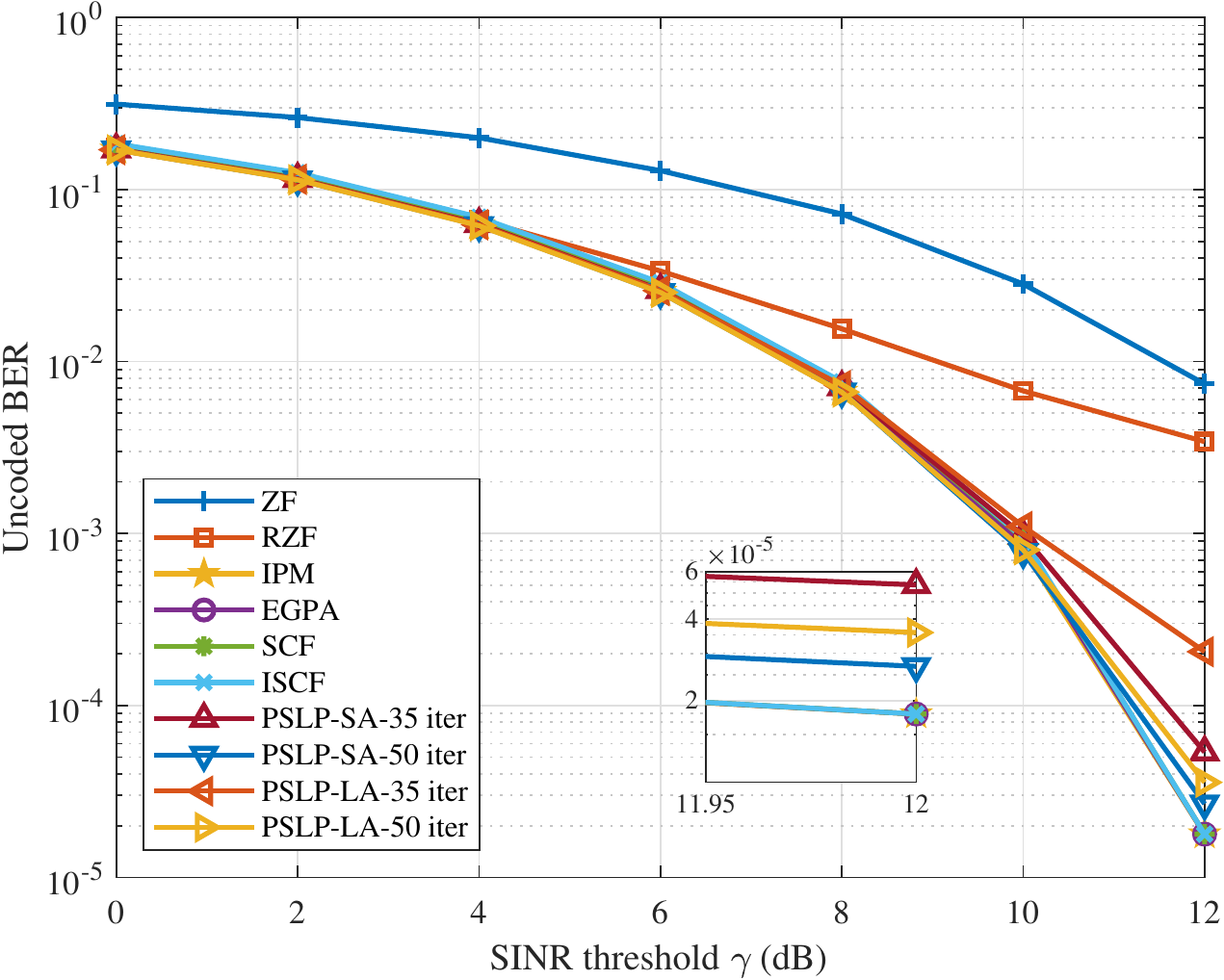}
\label{fig_second_case_ber}}
\hfil
\caption{Transmit power and uncoded BER performance of different schemes in different SINR thresholds, QPSK, $N_t=128$, $K=112$, $N=64$.}
\label{fig_performance}
\end{figure}

We compare the performance of the proposed approach and other schemes of interest in the view of transmit power and uncoded bit error rate (BER).

In Fig. \ref{fig_first_case_power} we depict the average transmit power with the SINR threshold for the same system setting.  The transmit power of the proposed parallel SLP algorithms approach those of the IPM from low to high, since we initialize the transmit signal as a zero vector. Specifically, the early termination of the PSLP-SA and PSLP-LA algorithm at 35 iterates leads to a suboptimal solution of nearly 0.15 dB and 0.1 dB transmit power gap, respectively. When the number of iterations of PSLP reaches 50, all the proposed schemes have optimal transmit power.

Fig. \ref{fig_second_case_ber} shows the uncoded BER performance of the proposed parallel SLP approach compared to other schemes at various SINR thresholds for the same system setting. At the interference-limited medium-to-high SINR threshold region, the SLP schemes implemented by the proposed parallel approach, IPM, SCF, ISCF, and EGPA all achieve lower uncoded BER over the ZF and RZF scheme. The performance of the proposed approach increases stably with the number of iterations, providing a performance-complexity trade-off. With sufficient iterates, the uncoded BER performance of the proposed approach matches that of the IPM.

\begin{figure}[!t]
\begin{minipage}{0.5\linewidth}
\centering
\includegraphics[width=18pc]{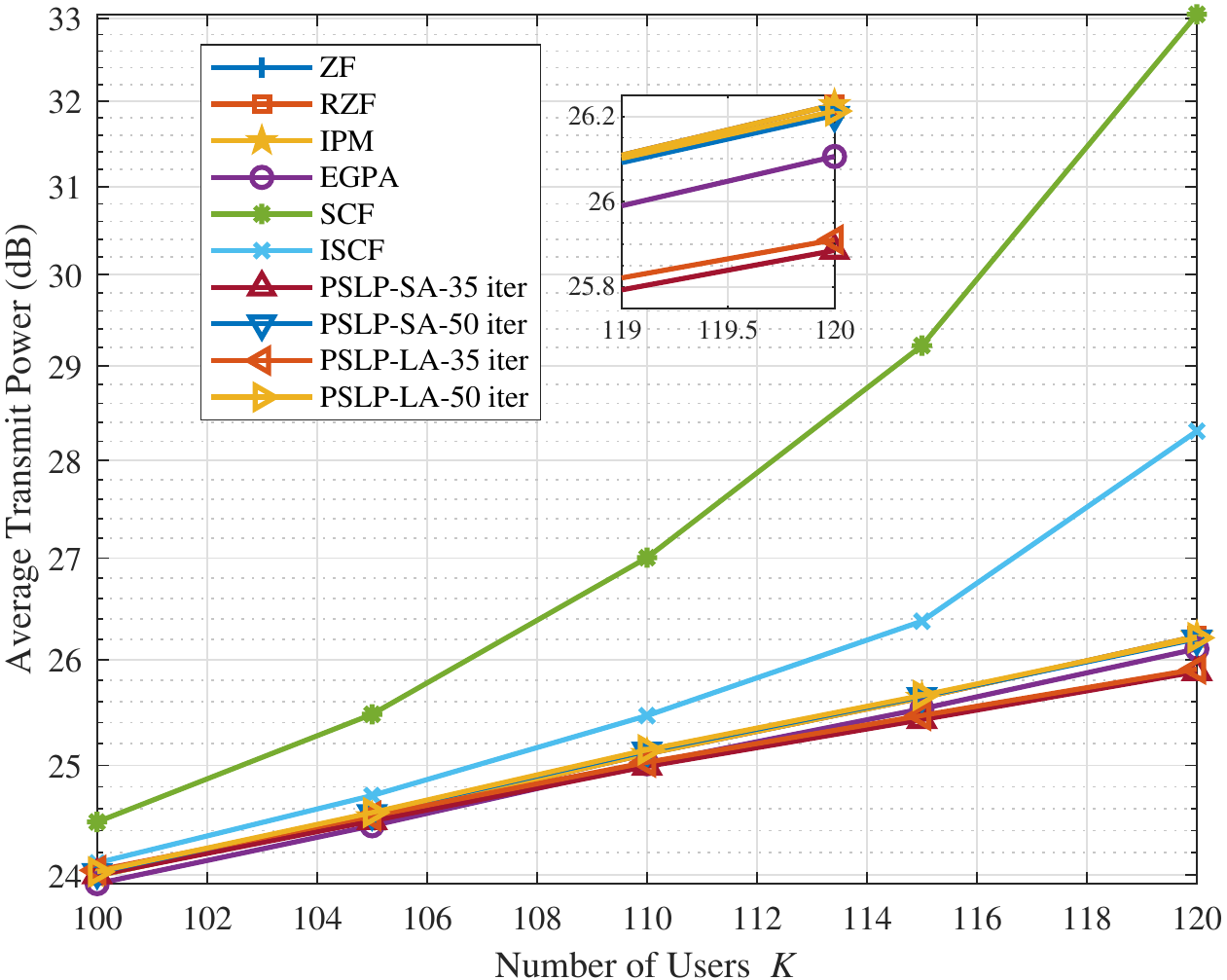}
\caption{Transmit power of different schemes in different number of users, QPSK, $N_t=128$, $\gamma=12$ dB,  $N=64$.}
\label{fig_performanceK}
\end{minipage}
\hfill
\begin{minipage}{0.5\linewidth}
\centering
\includegraphics[width=18pc]{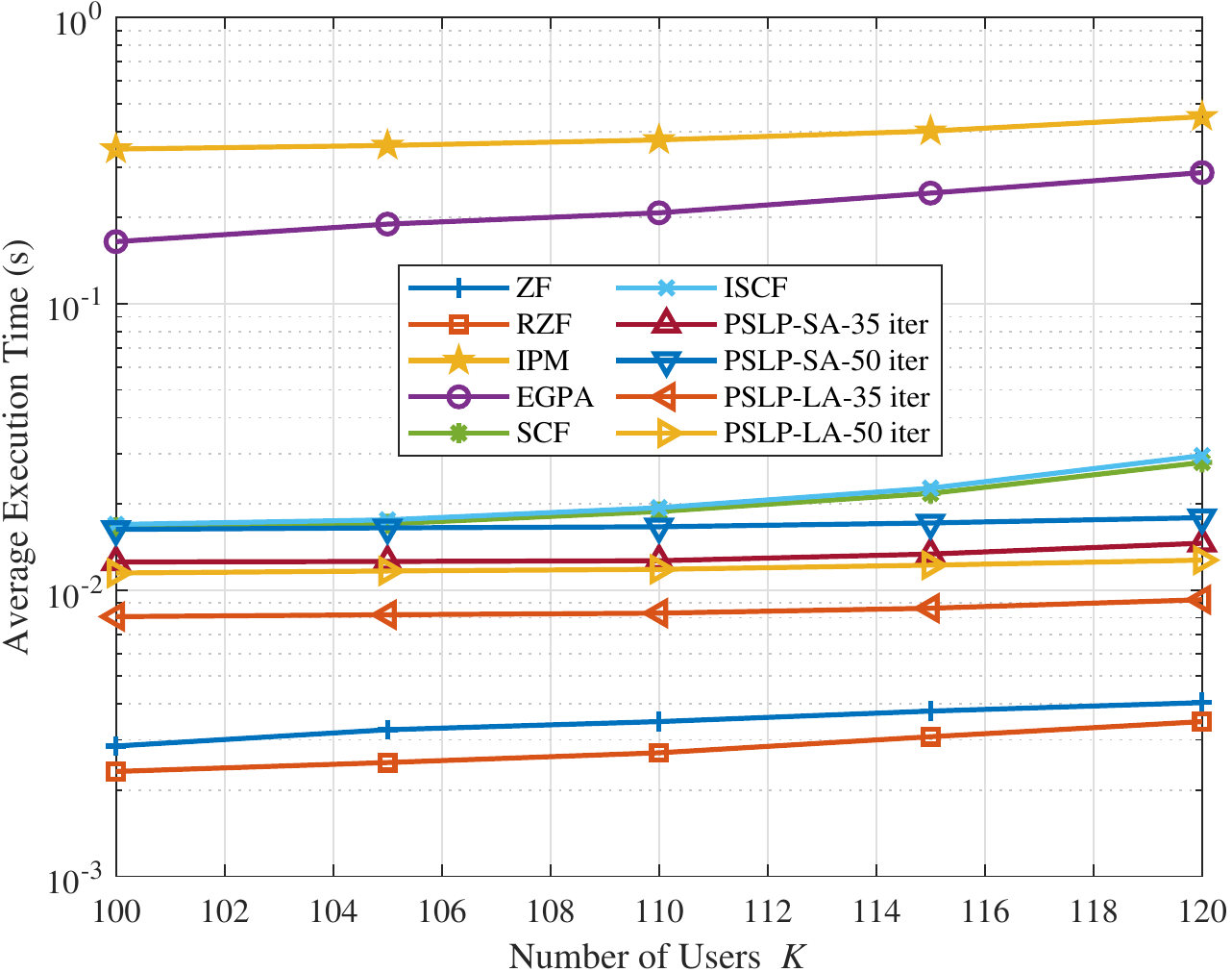}
\caption{Average execution time of different schemes in different number of users, QPSK, $N_t=128$, $\gamma=12$ dB,  $N=64$.}
\label{fig_timeK}
\end{minipage}
\end{figure}

We then demonstrate the effectiveness of the proposed parallel SLP approach under a varying number of users in Fig. \ref{fig_performanceK}. The system has 128 transmit antennas, and the number of users varies from 100 to 120. The SINR threshold is 12 dB. The trend can be seen in Fig. \ref{fig_performanceK} is that, unlike the ZF and RZF schemes as well as the SCF and ISCF schemes, whose transmit power performance severely deteriorates when the number of users increases, both the proposed PSLP-SA and PSLP-LA algorithms and EPGA show robustness among various load levels. Because the proposed parallel SLP approach and EPGA solve the PM-SLP problem successfully, regardless of the problem size. The performance of the early terminated parallel SLP approach at 35 iterations is degraded by the increasing system load, which validates that the larger problem size needs more iterations to converge.

\subsection{Computational Complexity Comparison}
Table \ref{table_time} presents the average number of iterations and execution time per channel realization of the proposed schemes, where `MI' and `SVD' denote the plain implementation with matrix inversion in each iteration in Section \ref{subsec_closed-form} and the SVD-based efficient implementation in Section \ref{efficientAlgr}, respectively. The stopping criterion is $\Delta\mathbf{x}^t < 10^{-3}$. We can observe that the proposed PIF-SLP algorithm with the adaptive tuning strategy (PSLP-LA) is the most efficient in reaching the given iterate gap.
\begin{table}[!htbp] 
\renewcommand{\arraystretch}{1.3}
\caption{Average Number of Iterations and Execution Time Over 2000 Random Channel Realizations, QPSK, $N_t=128$, $K = 112$,  $N=64$.}
\label{table_time}
\centering
\begin{tabular}{cccccccccc} 
\toprule 
\multicolumn{1}{c}{\multirow{2}{*}{Schemes}}& \multicolumn{2}{c}{$\gamma$ = 5 dB}& &\multicolumn{2}{c}{$\gamma$ = 10 dB}\\
\cline{2-6}
\multicolumn{1}{c}{}&Iterations&Time (s)&&Iterations&Time (s)\\  
\hline 
\multicolumn{1}{l}{PSLP-SA/MI}&62.3255&0.0401& &62.1500&0.0397\\   
\multicolumn{1}{l}{PSLP-SA/SVD}&62.3255&0.0242& &62.1500&0.0240\\
\multicolumn{1}{l}{PSLP-LA}&67.3320&0.0156& &67.1945&0.0156\\
\multicolumn{1}{l}{PSLP-SC/MI}&133.0500&0.0742& &133.3380&0.0741\\   
\multicolumn{1}{l}{PSLP-SC/SVD}&133.0500&0.0426& &133.3380&0.0425\\
\multicolumn{1}{l}{PSLP-LC}&126.1735&0.0293& &126.2845&0.0292\\
\bottomrule
\end{tabular}
\end{table}

Fig. \ref{fig_timeK} compares the average execution time required per channel realization of the concerned schemes. The system setting is the same as Fig. \ref{fig_performanceK}. We notice that implementing the parallel approach in physical parallel computing processors is beyond the range of this paper, thus the execution time for the proposed parallel SLP approach is the total time required for MATLAB simulation, which is an overestimate. It is observed that the proposed approach exhibits the lowest time complexity among other compared iterative schemes, i.e., IPM and EPGA. It also outperforms the closed-form solutions such as the SCF and the ISCF, excluding the heuristic ZF and RZF schemes.

\section{Conclusion}
\label{conclusion}
In this paper, parallel and decentralized processing for CI-based SLP is proposed for a massive MU-MISO downlink system based on ADMM. By reformulating the canonical PM-SLP optimization problem and introducing a slack variable vector, we transfer the original problem into separable equality constrained optimization, which is well-suited for the application of parallel processing. The augmented Lagrangian method is used to acquire an unconstrained problem formulation, which is further decomposed into several parallel subproblems via the PJ-ADMM framework. The sufficient condition for global convergence of the parallel CI-based SLP approach is derived, based on which a PIF-SLP algorithm is proposed to further alleviate the computational burden. Numerical results show the superiority of the proposed algorithms in terms of computational efficiency over state-of-the-art works, without compromising transmit power or BER performance.


%

\appendices
\section{Proof of Theorem 1}
\label{proof}
\begin{IEEEproof}[\bf Proof]
The first-order optimal condition for $\mathbf{x}_i$ is given by
\begin{IEEEeqnarray}{rCl}
\label{first-order condition}
\mathbf{A}^T_{i}\left(\boldsymbol{\lambda}^{t}-\rho\sum_{j=1}^{N}\mathbf{A}_{j}\mathbf{x}^{t}_j+\mathbf{b}+\mathbf{c}^{t+1}\right)+\mathbf{P}_i\left(\mathbf{x}^t_i-\mathbf{x}^{t+1}_i\right)+ \rho\mathbf{A}^T_{i}\mathbf{A}_{i}\left(\mathbf{x}^t_i-\mathbf{x}^{t+1}_i\right)\in\partial f_i\left(\mathbf{x}^{t+1}_i\right),\IEEEeqnarraynumspace
\end{IEEEeqnarray}
where $f_i\left(\mathbf{x}_i\right)\triangleq\left\|\mathbf{x}_i\right\|^2$.

Denoting $\hat{\boldsymbol{\lambda}}=\boldsymbol{\lambda}^{t}-\rho\left(\mathbf{A}\mathbf{x}^{t+1}-\mathbf{b}-\mathbf{c}^{t+1}\right)$, then the first-order optimal condition (\ref{first-order condition}) turns to
\begin{IEEEeqnarray}{rCl}
\mathbf{A}^T_{i}\left[\hat{\boldsymbol{\lambda}}-\rho\sum_{j=1}^{N}\mathbf{A}_{j}\left(\mathbf{x}^{t}_j-\mathbf{x}^{t+1}_j\right)\right]+\left(\mathbf{P}_i+\rho\mathbf{A}^T_{i}\mathbf{A}_{i}\right)\left(\mathbf{x}^t_i-\mathbf{x}^{t+1}_i\right) &\in &\partial f_i\left(\mathbf{x}^{t+1}_i\right).\IEEEeqnarraynumspace
\end{IEEEeqnarray}

Assum there exist a saddle point $\mathbf{u}^{*}=\left(\mathbf{x}^{*}_1,\mathbf{x}^{*}_2\cdots,\mathbf{x}^{*}_N,\boldsymbol{\lambda}^{*}\right)$ for PM-SLP. From the convexity of $f_i\left(\mathbf{x}_i\right)$, we have
\begin{IEEEeqnarray}{rCl}
\label{inequality of convexity}
\left(\partial f_i\left(\mathbf{x}^{t+1}_i\right)-\partial f_i\left(\mathbf{x}^*_i\right)\right)^T\left(\mathbf{x}^{t+1}_i-\mathbf{x}^*_i\right)\geq 0.
\end{IEEEeqnarray}
The stationarity condition of KKT conditions is given by
\begin{IEEEeqnarray}{rCl}
& &\mathbf{A}^T_{i}\boldsymbol{\lambda}^*\in\partial f_i\left(\mathbf{x}^*_i\right).
\end{IEEEeqnarray}
Thus (\ref{inequality of convexity}) can be written as
\begin{IEEEeqnarray}{rCl}
\IEEEeqnarraymulticol{3}{l}{\left[ \mathbf{A}^T_{i}\left( \hat{\boldsymbol{\lambda}}-\boldsymbol{\lambda}^*-\rho\sum_{j=1}^{N}\mathbf{A}_{j}\left( \mathbf{x}^{t}_j-\mathbf{x}^{t+1}_j\right) \right)\right]^T
\left(\mathbf{x}^{t+1}_i-\mathbf{x}^*_i\right)}\nonumber\\ \qquad \qquad \qquad \qquad \qquad \qquad+\left(\mathbf{x}^{t+1}_i-\mathbf{x}^*_i\right)^T\left(\mathbf{P}_i+\rho\mathbf{A}^T_i\mathbf{A}_i\right)\left( \mathbf{x}^t_i-\mathbf{x}^{t+1}_i\right) \geq 0.\IEEEeqnarraynumspace
\end{IEEEeqnarray}
Summing the above inequality over all $i$, we obtain
\begin{IEEEeqnarray}{rCl}
\left( \hat{\boldsymbol{\lambda}}-\boldsymbol{\lambda}^*\right)^T\mathbf{A}\left(\mathbf{x}^{t+1}-\mathbf{x}^*\right)+\sum^N_{i=1}\left(\mathbf{x}^{t+1}_i-\mathbf{x}^*_i\right)^T\left(\mathbf{P}_i+\rho\mathbf{A}^T_{i}\mathbf{A}_{i}\right)\left( \mathbf{x}^t_i-\mathbf{x}^{t+1}_i\right)\nonumber\\
\qquad \qquad\geq  \rho\left( \mathbf{x}^{t}-\mathbf{x}^{t+1}\right)^T\mathbf{A}^T\mathbf{A}\left(\mathbf{x}^{t+1}-\mathbf{x}^*\right).\IEEEeqnarraynumspace
\label{aggregated inequality}
\end{IEEEeqnarray}

Note that
\begin{IEEEeqnarray}{C}
\label{dualUpdateEqua}
\mathbf{A}\left(\mathbf{x}^{t+1}-\mathbf{x}^*\right)=\frac{1}{\beta\rho}\left(\boldsymbol{\lambda}^t-\boldsymbol{\lambda}^{t+1}\right),\\
\label{dualInsertEqua}
\hat{\boldsymbol{\lambda}}-\boldsymbol{\lambda}^*=\left(\hat{\boldsymbol{\lambda}}-\boldsymbol{\lambda}^{t+1}\right)+\left(\boldsymbol{\lambda}^{t+1}-\boldsymbol{\lambda^*}\right)
=\frac{\beta-1}{\beta}\left(\boldsymbol{\lambda}^t-\boldsymbol{\lambda}^{t+1}\right)+\left(\boldsymbol{\lambda}^{t+1}-\boldsymbol{\lambda}^*\right).
\IEEEeqnarraynumspace
\end{IEEEeqnarray}
Substituting (\ref{dualUpdateEqua}) and (\ref{dualInsertEqua}) into (\ref{aggregated inequality}), we obtain
\begin{IEEEeqnarray}{rCl}
\label{primal dual inequality}
\left( \boldsymbol{\lambda}^{t+1}-\boldsymbol{\lambda}^*\right)^T\frac{1}{\beta\rho}\left(\boldsymbol{\lambda}^t-\boldsymbol{\lambda}^{t+1}\right)+\sum^N_{i=1}\left(\mathbf{x}^{t+1}_i-\mathbf{x}^*_i\right)^T\left(\mathbf{P}_i+\rho\mathbf{A}^T_{i}\mathbf{A}_{i}\right)\left( \mathbf{x}^t_i-\mathbf{x}^{t+1}_i\right)\nonumber\\\geq \frac{1-\beta}{\beta^2\rho}\left\|\boldsymbol{\lambda}^t-\boldsymbol{\lambda}^{t+1}\right\|^2+ \frac{1}{\beta}\left(\boldsymbol{\lambda}^t-\boldsymbol{\lambda}^{t+1}\right)^T\mathbf{A}\left(\mathbf{x}^{t}-\mathbf{x}^{t+1}\right).\IEEEeqnarraynumspace
\end{IEEEeqnarray}

For notation simplicity, denoting $\mathbf{G}_x\triangleq{\left[\begin{IEEEeqnarraybox*}[][c]{,c/c/c,}
\mathbf{P}_1+\rho\mathbf{A}^T_1\mathbf{A}_1 & & \\ 
 & \ddots & \\
 &&\mathbf{P}_N+\rho\mathbf{A}^T_N\mathbf{A}_N \end{IEEEeqnarraybox*}\right]}$, $\mathbf{G}\triangleq{\left[\begin{IEEEeqnarraybox*}[][c]{,c/c,}
\mathbf{G}_x &  \\ 
 &\frac{1}{\beta\rho}\mathbf{I} \end{IEEEeqnarraybox*}\right]}$, $\mathbf{Q}\triangleq{\left[\begin{IEEEeqnarraybox*}[][c]{,c/c/c/c,}
\mathbf{P}_1+\rho\mathbf{A}^T_1\mathbf{A}_1 & & &\frac{1}{\beta}\mathbf{A}^T_1 \\ 
 & \ddots & & \vdots\\
 &&\mathbf{P}_N+\rho\mathbf{A}^T_N\mathbf{A}_N & \frac{1}{\beta}\mathbf{A}^T_1\\
\frac{1}{\beta}\mathbf{A}^T_1 &\cdots&\frac{1}{\beta}\mathbf{A}^T_N&\frac{2-\beta}{\rho\beta^2}\mathbf{I}\end{IEEEeqnarraybox*}\right]}.$
Essentially, from (\ref{primal dual inequality}) we have
\begin{IEEEeqnarray}{rCl}
\left(\mathbf{u}^t-\mathbf{u}^{t+1}\right)^T\mathbf{G}\left(\mathbf{u}^{t+1}-\mathbf{u}^{*}\right)
\geq \frac{1-\beta}{\beta^2\rho}\left\|\boldsymbol{\lambda}^t-\boldsymbol{\lambda}^{t+1}\right\|^2+\frac{1}{\beta}\left(\boldsymbol{\lambda}^t-\boldsymbol{\lambda}^{t+1}\right)^T\mathbf{A}\left(\mathbf{x}^{t}-\mathbf{x}^{t+1}\right).\IEEEeqnarraynumspace
\end{IEEEeqnarray}
Thus we have the relationship
\begin{IEEEeqnarray}{rCl}
\label{GQinequa}
\IEEEyesnumber\IEEEyessubnumber*
\left\|\mathbf{u}^t-\mathbf{u}^*\right\|^2_{\mathbf{G}}-\left\|\mathbf{u}^{t+1}-\mathbf{u}^*\right\|^2_{\mathbf{G}}&=&2\left(\mathbf{u}^t-\mathbf{u}^{t+1}\right)^T\mathbf{G}\left(\mathbf{u}^{t+1}-\mathbf{u}^*\right)+\left\|\mathbf{u}^t-\mathbf{u}^{t+1}\right\|^2_{\mathbf{G}}\\
&\geq & \frac{2-2\beta}{\beta^2\rho}\left\|\boldsymbol{\lambda}^t-\boldsymbol{\lambda}^{t+1}\right\|^2+\frac{2}{\beta}\left(\boldsymbol{\lambda}^t-\boldsymbol{\lambda}^{t+1}\right)^T\mathbf{A}\left(\mathbf{x}^{t}-\mathbf{x}^{t+1}\right)\nonumber\\&&+\left\|\mathbf{u}^t-\mathbf{u}^{t+1}\right\|^2_{\mathbf{G}}
\\&=&\frac{2-2\beta}{\beta^2\rho}\left\|\boldsymbol{\lambda}^t-\boldsymbol{\lambda}^{t+1}\right\|^2+\frac{2}{\beta}\left(\boldsymbol{\lambda}^t-\boldsymbol{\lambda}^{t+1}\right)^T\mathbf{A}\left(\mathbf{x}^{t}-\mathbf{x}^{t+1}\right)\nonumber\\&&+\left\|\mathbf{x}^t-\mathbf{x}^{t+1}\right\|^2_{\mathbf{G}_\mathbf{x}}+\frac{1}{\beta\rho}\left\|\boldsymbol{\lambda}^t-\boldsymbol{\lambda}^{t+1}\right\|^2
\\ &=&\frac{2-\beta}{\beta^2\rho}\left\|\boldsymbol{\lambda}^t-\boldsymbol{\lambda}^{t+1}\right\|^2+\frac{2}{\beta}\left(\boldsymbol{\lambda}^t-\boldsymbol{\lambda}^{t+1}\right)^T\mathbf{A}\left(\mathbf{x}^{t}-\mathbf{x}^{t+1}\right)\nonumber\\&&+\left\|\mathbf{x}^t-\mathbf{x}^{t+1}\right\|^2_{\mathbf{G}_\mathbf{x}}
\\&=&\left\|\mathbf{u}^t-\mathbf{u}^{t+1}\right\|^2_{\mathbf{Q}}.
\end{IEEEeqnarray}
To prove the convergence of the PJ-ADMM for PM-SLP is reduced to ensure that $\mathbf{Q}$ is positive semi-definite. For any $\mathbf{u}\in\mathbb{R}^{2N_t+2K}$, we have
\begin{IEEEeqnarray}{rCl}
\left\|\mathbf{u}\right\|^2_{\mathbf{Q}}=\left\|\mathbf{x}\right\|^2_{\mathbf{G}_\mathbf{x}}+\frac{2-\beta}{\beta^2\rho}\left\|\boldsymbol{\lambda}\right\|^2+\frac{2}{\beta}\boldsymbol{\lambda}^T\mathbf{A}\mathbf{x}.
\end{IEEEeqnarray}
Using the basic inequality
\begin{IEEEeqnarray}{rCl}
\label{basic inequality}
\frac{2}{\beta}\boldsymbol{\lambda}^T\mathbf{A}\mathbf{x}&=&\sum^N_{i=1}\frac{2}{\beta}\boldsymbol{\lambda}^T\mathbf{A}_i\mathbf{x}_i\geq  -\sum^N_{i=1}\left(\frac{\epsilon_i}{\rho\beta^2}\left\|\boldsymbol{\lambda}\right\|^2+\frac{\rho}{\epsilon_i}\left\|\mathbf{A}_i\mathbf{x}_i\right\|^2\right),
\end{IEEEeqnarray}
for any $\epsilon_i>0$, we have
\begin{IEEEeqnarray}{rCl}
\left\|\mathbf{u}\right\|^2_{\mathbf{Q}}&\geq &\sum^N_{i=1}\left\|\mathbf{x}_i\right\|^2_{\mathbf{P}_i+\rho\mathbf{A}^T_i\mathbf{A}_i-\frac{\rho}{\epsilon_i}\mathbf{A}^T_i\mathbf{A}_i}+\frac{2-\beta-\sum^N_{i=1}\epsilon_i}{\beta^2\rho}\left\|\boldsymbol{\lambda}\right\|^2.
\end{IEEEeqnarray}
Therefore, $\mathbf{Q}$ is positive semi-definite if
\begin{IEEEeqnarray}{rCl}
\setcounter{equation}{28}
\mathbf{P}_i&\succeq&\rho(\frac{1}{\epsilon_i}-1)\mathbf{A}^T_i\mathbf{A}_i, \forall i, \sum^N_{i=1}\epsilon_i\leq2-\beta,
\end{IEEEeqnarray}
where $\epsilon_i>0$.

If the sufficient condition is satisfied, then the error metric $\left\|\mathbf{u}^t-\mathbf{u}^*\right\|^2_{\mathbf{G}}$ is monotonically non-decreasing, and the sequence $\{{\mathbf{u}}^t\}$ generated by the PJ-ADMM is contractive. The global convergence of the algorithm follows immediately from the analysis of the contraction method \cite{he1997class}.
\end{IEEEproof}


%
%

\ifCLASSOPTIONcaptionsoff
  \newpage
\fi



%
%
%

\bibliographystyle{IEEEtran}
\bibliography{IEEEabrv,references}

%

%
%
%




\end{document}